\documentstyle[12pt,psfig]{article}

\def\ltap{\raisebox{-.4ex}{\rlap{$\sim$}} \raisebox{.4ex}{$<$}}
\def\gtap{\raisebox{-.4ex}{\rlap{$\sim$}} \raisebox{.4ex}{$>$}}

\begin{document}
\begin{titlepage}
\today          \hfill
\begin{center}
\hfill    LBL-40121 \\

\vskip .25in
{\large \bf Improving the Fine Tuning in 
Models of Low Energy Gauge Mediated Supersymmetry Breaking 
}
\footnote{This work was supported in part
by the Director, Office of Energy
Research, Office of High Energy and Nuclear Physics, 
Division of High
Energy Physics of the U.S. Department of Energy 
under Contract
DE-AC03-76SF00098 and in part by the National Science 
Foundation
under grant PHY-90-21139.}

\vskip .25in

\vskip .25in
K. Agashe\footnote{email: agashe@theor3.lbl.gov.} and M. Graesser
\footnote{email: mlgraesser@lbl.gov}

{\em Theoretical Physics Group\\
    Ernest Orlando Lawrence Berkeley National Laboratory\\
      University of California\\
    Berkeley, California 94720}
\end{center}

\vskip .25in

\begin{abstract}
The 
fine tuning
in models of low energy 
gauge mediated supersymmetry breaking 
required 
to obtain the correct $Z$ mass is quantified.  To 
alleviate the fine tuning problem,
a model with split $(5+\bar{5})$
messenger fields
is presented.
This model has additional triplets in the low 
energy theory which get a mass of $O(500)$ GeV
from a coupling to a singlet.  
The improvement in fine tuning is quantified
and the spectrum in this model is discussed. 
The same model with the above singlet coupled to the 
Higgs doublets 
to generate the $\mu$ term is also discussed. 
A Grand Unified version of the model is constructed and 
a known doublet-triplet splitting mechanism is used
to split the messenger $(5+\bar{5})$'s.
A complete 
model is presented and some 
phenomenological constraints are discussed.
\end{abstract}
\end{titlepage}
\renewcommand{\thepage}{\roman{page}}
\setcounter{page}{2}
\mbox{ }

\vskip 1in

\begin{center}
{\bf Disclaimer}
\end{center}

\vskip .2in

\begin{scriptsize}
\begin{quotation}
This document was prepared as an account of work sponsored by the 
United
States Government. While this document is believed to contain 
correct
 information, neither the United States Government nor any agency
thereof, nor The Regents of the University of California, nor any 
of their
employees, makes any warranty, express or implied, or assumes any legal
liability or responsibility for the accuracy, completeness, 
or usefulness
of any information, apparatus, product, or process disclosed, or 
represents
that its use would not infringe privately owned rights.  
Reference herein
to any specific commercial products process, or service by 
its trade name,
trademark, manufacturer, or otherwise, does not necessarily 
constitute or
imply its endorsement, recommendation, or favoring by the 
United States
Government or any agency thereof, or The Regents of the 
University of
California.  The views and opinions of authors expressed herein 
do not
necessarily state or reflect those of the United States 
Government or any
agency thereof, or The Regents of the University of California.
\end{quotation}
\end{scriptsize}

\vskip 2in
\begin{center}
\begin{small}
{\it Lawrence Berkeley Laboratory is an equal opportunity employer.}
\end{small}
\end{center}

\newpage
\renewcommand{\thepage}{\arabic{page}}
\setcounter{page}{1}

\section{Introduction}

One of the outstanding problems 
of particle physics is the origin
of electroweak symmetry breaking (EWSB).
In the Standard Model (SM), this is 
achieved by one Higgs doublet which acquires 
a vacuum expectation value (vev) due 
to a negative mass squared which is put in by hand.
The SM has the well known gauge hierarchy problem 
\cite{susskind}. 
It is known that 
supersymmetry (SUSY) \cite{susy} stabilises 
the hierachy between the weak scale and 
some other high scale without any fine tuning if
the masses of the superpartners are less than 
few TeV
\cite{barbieri1,anderson}. 
The Minimal Supersymmetric Standard Model 
(MSSM) is considered as a low energy 
effective theory in which the soft SUSY 
breaking terms 
at the correct scale
are put in by hand. This raises 
the question : what is the origin of 
these soft mass terms, {\it i.e.}, how is 
SUSY broken ? 
If SUSY is broken spontaneously at tree
level in the MSSM, then there is a 
colored scalar 
lighter than the up or down quarks \cite{georgi}.
So, the superpartners have to 
acquire mass through 
radiative corrections. Thus,
we need a ``hidden'' sector where 
SUSY is broken spontaneously at tree level and then
communicated to the MSSM by some 
``messengers''. 

There are two problems here: how
is SUSY broken in the hidden 
sector at the right scale  and what are the messengers ? 
There are models in which a 
dynamical superpotential 
is generated by non-perturbative effects
which breaks SUSY \cite{affleck}. The SUSY 
breaking scale is related to the Planck scale by
dimensional transmutation.
 Two possibilities have been 
discussed in the literature for
the messengers. One is gravity 
which couples to both the sectors \cite{hall}. 
In a supergravity
theory, there are non-renormalizable 
couplings between the two sectors which generate 
soft SUSY breaking operators in the 
MSSM  
once SUSY is broken in the ``hidden'' sector. 
In the absence of a flavor symmetry, this theory
has to be fine tuned to give  
almost degenerate squarks and sleptons of the first two 
generations which
is required by Flavor Changing 
Neutral Current (FCNC) phenomenology
 \cite{georgi,gabbiani}. The other messengers
are the SM gauge interactions \cite{wise}. In 
these models, the scalars of the first two generations
are naturally degenerate since they 
have the same gauge quantum numbers. This is 
an attractive feature of these models, 
since 
the FCNC
constraints are naturally avoided and no
fine tuning between the masses of the first two 
generation scalars is required. If this lack 
of fine tuning is a compelling argument in favour 
of these models, then it   
is important to 
investigate
whether other sectors of these models are fine tuned.
In fact, we will argue (and this is also discussed in
\cite{nelson2,arkani,strumia}) that the
 minimal model (to be defined in section 2) of low energy 
gauge mediated SUSY breaking  
requires a minimum
7$\%$ fine tuning to generate a correct vacuum ($Z$ mass).
Further, if a gauge-singlet is introduced to generate 
the ``$\mu$'' and ``$B\mu$'' terms, then the minimal model 
of low energy
gauge mediated SUSY breaking
requires a minimum 1$\%$ fine tuning to 
correctly break the electroweak symmetry.
These fine tunings
makes it difficult to understand, within the 
context of these models, 
how SUSY is to offer
some understanding of the origin of electroweak symmetry
breaking and the scale of the $Z$ and $W$ gauge boson
masses.

Our paper is organized as follows. 
In section \ref{mess}, we briefly review
both the ``messenger sector'' in low energy gauge 
mediated SUSY breaking 
models that communicates SUSY 
breaking to the Standard Model and the pattern of the
sfermion and gaugino masses that follow. 
Section \ref{finetune} quantifies
the fine tuning in the minimal model using the 
Barbieri-Giudice 
criterion \cite{barbieri1}.
We show that a fine tuning of $\approx 7 \%$ is 
required in the Higgs sector to obtain $m_Z$.
Section \ref{toymodel} describes 
a toy model 
with split $(5+\bar{5})$ messenger representations 
that 
improves the fine tuning. To maintain gauge coupling 
unification,
additional triplets are added to the 
low energy theory. They
acquire a mass of $O(500)$ GeV
by a coupling to a singlet. 
The fine tuning in this model 
is 
improved to $\sim 40 \%$.
The sparticle phenomenology of these models is also 
discussed.
In section \ref{NMSSM}, 
we discuss a version of the
toy model where the above mentioned
singlet 
generates the $\mu$ and $\mu^2_3$ terms. 
This is identical to the Next-to-Minimal 
Supersymmetric Standard Model (NMSSM) \cite{fayet} 
with a particular pattern for the 
soft SUSY breaking operators 
that follows from gauge mediated 
SUSY breaking and our solution to
the fine tuning problem.
We show that this model is tuned to $\sim 20 \%$, even if
LEP does not discover SUSY/light Higgs.
We also show that the NMSSM with one complete messenger 
$(5 + \bar{5})$
is fine tuned to $\sim 1 \%$. 
We  
discuss, in section \ref{GUT}, how 
it is possible to make our toy model compatible with
a Grand Unified Theory (GUT) \cite{splitting}
based upon the gauge group $SU(5) \times SU(5)$.
The doublet-triplet splitting mechanism 
of Barbieri, Dvali and Strumia
\cite{barbieri2} is used to 
split both the messenger representations and the Higgs 
multiplets. 
In 
section 7, we present a model in which all 
operators consistent with symmetries are present and 
demonstrate that the low energy theory is the 
model of section \ref{NMSSM}. 
In this model $R$-parity $(R_p)$ is the unbroken 
subgroup of a $Z_4$ global discrete symmetry that is
required to solve the doublet-triplet splitting problem.
Our model has some 
metastable particles which might cause a cosmological
problem. In the appendix, we give the expressions 
for the Barbieri-Giudice parameters
(for the fine tuning) for the MSSM and the NMSSM.

\section{Messenger Sector}
\label{mess}
In the models of low energy gauge mediated SUSY breaking
\cite{nelson2,nelson} (henceforth called LEGM models), SUSY
breaking occurs dynamically in a ``hidden'' sector of 
the theory
at a scale $\Lambda_{dyn}$ that is generated through
dimensional transmutation. SUSY breaking is
communicated to the Standard Model fields in 
two stages.
First, a non-anomalous 
$U(1)$ global symmetry of the hidden sector
is weakly gauged. This $U(1)_X$ gauge interaction
communicates SUSY breaking from the original 
SUSY
breaking sector to a messenger sector at a scale
$\Lambda_{mess}\sim \alpha_X \Lambda_{dyn}/(4\pi)$ as follows.
The particle content
in the messenger sector consists of fields $\phi_{+}$, 
$\phi_{-}$
charged under this $U(1)_X$, a gauge singlet field $S$,
and vector-like
fields that carry Standard Model quantum numbers 
(henceforth called 
messenger quarks and leptons).
In the minimal LEGM model,
there is one set of vector-like fields, $\bar{q}$, $l$,
and $q$, $\bar{l}$  that together
form a $(\bar{5} + 5)$ of $SU(5)$.
This is a suffucient condition to maintain unification of
the SM gauge couplings. The superpotential
in the minimal model is
\begin{equation}
W_{mess}=\lambda _{\phi} \phi_{+} \phi_{-}S +
\frac{1}{3} \lambda _S S^3
+\lambda_q S q\bar{q}+
\lambda_l S l\bar{l}.
\label{eq:potential}
\end{equation}
The scalar potential is
\begin{equation}
V=\sum_i|F_i|^2+m^2_+|\phi_+|^2+m^2_-|\phi_-|^2.
\end{equation}
In the models of \cite{nelson2,nelson}, 
the $\phi _+, \phi _-$ fields 
communicate (at two loops) with the hidden sector fields 
through the 
$U(1)$ gauge interactions. 
Then, SUSY breaking
in the original
sector generates a negative value 
$ \sim - \left( \alpha_X \Lambda_{dyn}
\right)^2/(4\pi)^2$
for the mass
parameters $m^2_+$,
$m^2_-$ of the
$\phi_+$ and $\phi_-$ fields. This drives vevs
of $O \left( \Lambda _{mess} \right)$
for
the scalar components of both
$\phi_+$ and $\phi_-$,
and also for the scalar and $F$-component of $S$
if the couplings
$\lambda_S$, $g_X$ and $\lambda _{\phi}$ satisfy the 
inequalities
derived in
\cite{arkani,randall}.\footnote{
This point in field space is a local
minimum. There is a deeper minimum
where SM is broken \cite{arkani,randall}. To avoid this 
problem,
we can, for example, add another singlet to the 
messenger sector \cite{arkani}.
This does not change our conclusions about the fine tuning.}
Generating a vev for both the scalar and $F$-component
of $S$ is
crucial, since this generates a non-supersymmetric
spectrum for the
vector-like fields $q$ and $l$.
The spectrum of each vector-like messenger field
consists of
two complex scalars
with masses $M^2 \pm B$ and
two Weyl fermions with mass $M$
where $M=\lambda S$, $B=\lambda F_S$
and $\lambda$ is the coupling of the vector-like fields to
$S$. Since we do not want the SM to be broken at this
stage, $M^2-B\ge$0.
In the second stage,
the messenger fields are
integrated out.
As these messenger fields have
SM gauge interactions,
SM
gauginos acquire masses at one loop
and the sfermions and Higgs acquire soft scalar
masses at two
loops \cite{wise}.
The gaugino masses at the scale at which the
messenger fields are integrated out, $\Lambda_{mess}
\approx M$ are \cite{nelson}
\begin{equation}
M_G =\frac{\alpha_G(\Lambda_{mess})}{4\pi}\Lambda_{SUSY}
\sum_m N^G_R(m)f_1\left(\frac{F_S}{\lambda_mS^2}\right).
\label{gaugino}
\end{equation}
The sum in equation \ref{gaugino} is over
messenger fields $(m)$ 
with normalization \\$ \hbox{Tr} (T^a T^b) = N^G_R(m)
\delta ^{ab}$ where the $T$'s are the generators 
of the gauge group $G$
in the representation $R$,
$f_1 (x)=1+O(x)$, and
$\Lambda_{SUSY}\equiv B/M=F_S/S=x\Lambda_{mess}$
with $x=B/M^2$.
\footnote{If all the dimensionless couplings in the 
superpotential are
of $O(1)$, then $x$ cannot be much smaller than 1.}
Henceforth, we will 
set $\Lambda_{SUSY}\approx\Lambda_{mess}$. The exact one loop 
calculation \cite{unpublished} of the gaugino mass shows that
$f_1(x) \leq$ 1.3 for $x \leq$ 1. 
The soft scalar masses at $\Lambda_{mess}$ are \cite{nelson}
\begin{equation}
{m_i}^2=2 \Lambda^2_{SUSY}\sum_{m,G}N^G_R(m)C^G_R(s_i)
\left(\frac{\alpha_G(\Lambda_{mess})}{4\pi}\right)^2 
f_2\left(\frac{F_S}
{\lambda_mS^2}\right),
\label{scalarmass}
\end{equation}
where
$C^G_R(s_i)$ is the Caismir of the representation of the 
scalar $i$
in the gauge group $G$ and $f_2 (x) = 1 + O(x)$.
The exact two loop calculation \cite{unpublished} 
which determines $f_2$ shows that for 
$x\leq$0.8 (0.9), $f_2$ differs from one by less 
than 1$\%$(5$\%$). 
Henceforth we shall 
use $f_1(x)=1$ and $f_2(x)=1$.
In the minimal 
LEGM model
\begin{equation}
M_G(\Lambda_{mess})=\frac{\alpha_G(\Lambda_{mess})}{4\pi}
\Lambda_{mess},
\end{equation}
\begin{eqnarray}
m^2(\Lambda_{mess})&=&2\Lambda^2_{mess}\times \\
 & & \left(C_3\left(
\frac{\alpha_3(\Lambda_{mess})}{4\pi}\right)^2  
  +  C_2\left(\frac{\alpha_2(\Lambda_{mess})}{4\pi}\right)^2+
\frac{3}{5}\left(\frac{\alpha_1(\Lambda_{mess})Y}{4\pi}
\right)^2\right),
\nonumber
\end{eqnarray}
where $Q=T_{3L}+Y$ and $\alpha_1$ is the $SU(5)$ normalized 
hypercharge coupling. Further,
$C_3=4/3$ and $C_2=3/4$ for colored 
triplets and electroweak doublets respectively. 

The spectrum in the models is determined by only a few 
unknown parameters. 
As equations 3 and 4
indicate, 
the SUSY 
breaking mass parameters for the Higgs, sfermions and
gauginos are
\begin{equation}
m_{\tilde{q}},m_{\tilde{g}}:m_{\tilde{L}},m_{H_i},
m_{\tilde{W}}:m_{\tilde{e}_R},m_{\tilde{B}}
\sim \alpha_3:\alpha_2:\alpha_1.
\end{equation}
The scale of $\Lambda_{mess}$ is chosen
to be $\sim$ 100 TeV so that the lightest 
of these particles escapes detection. The phenomenology
of the minimal LEGM model is discussed in detail in 
\cite{thomas}.

\section{Fine Tuning in the Minimal LEGM}
\label{finetune}

A desirable feature of gauge mediated SUSY breaking is
the natural suppression of FCNC
processes since the scalars with the same gauge
quantum numbers are degenerate \cite{wise}. 
But, the minimal LEGM model 
introduces a fine tuning in the Higgs sector 
unless the messenger scale is low.
This has been previously discussed in \cite{nelson2,arkani} 
and quantified 
more recently in \cite{strumia}. 
We outline the discussion in order to introduce some 
notation.

The superpotential for the MSSM is
\begin{equation}
W=\mu H_u H_d+W_{Yukawa}.
\end{equation}
The scalar potential is
\begin{equation}
V=\mu^2_1|H_u|^2+\mu^2_2|H_d|^2-(\mu^2_3H_u H_d + h.c.) 
\hbox{+D-terms}
+V_{1-loop},
\label{potential}
\end{equation}
where $V_{1-loop}$ is the one loop effective potential.
The vev of $H_u$ ($H_d$), denoted by $v_u (v_d)$,
 is responsible for giving mass
to the up (down)-type quarks, $\mu^2_1=m^2_{H_d}+{\mu}^2$,
$\mu^2_2=m^2_{H_u}+{\mu}^2$ and 
$\mu^2_3$,
\footnote{$\mu^2_3$ is often written as
$B\mu$.} $m^2_{H_u}$, $m^2_{H_d}$ are the SUSY 
breaking mass terms for the Higgs fields.
\footnote{The scale dependence
of the parameters appearing in the potential is implicit.} 
Extremizing this
potential determines, with $\tan\beta\equiv v_u/v_d$,
\begin{equation}
\frac{1}{2}{m_Z}^2=\frac{\tilde{\mu}^2_1-\tilde{\mu}^2_2
\tan^2\beta}{\tan^2\beta-1},
\label{mssm1}
\end{equation}
\begin{equation}
\sin 2\beta=2\frac{\mu^2_3}{\tilde{\mu}^2_1+\tilde{\mu}^2_2},
\label{mssm2}
\end{equation} 
where $\tilde{\mu}^2_i=\mu^2_i
+2\partial V_{1-loop}/\partial v^2_i$. For large 
$\tan\beta$, $m^2_Z/2\approx -(m^2_{H_u}+\mu^2)$. This 
indicates that if $|m^2_{H_u}|$ is large 
relative to $m^2_Z$, the $\mu^2$ term must cancel this
large number to reproduce the correct value for $m^2_Z$. 
This
introduces a fine tuning in the Higgs potential, that is
naively of the order
$m^2_Z/(2|m^2_{H_u}|)$. We shall show 
that this occurs in the minimal LEGM model. 

In the minimal LEGM model, a specification of the
messenger particle content and the 
messenger scale $\Lambda_{mess}$
fixes the sfermion and gaugino spectrum 
at that scale. For example, 
the soft scalar masses
for the Higgs fields are
$\approx \alpha_2 (\Lambda_{mess}) \Lambda_{mess}
/(4 \pi)$.
Renormalization
Group (RG) evolution from $\Lambda_{mess}$ to the 
electroweak scale reduces
$|m^2_{H_u}|$ due to the
large top quark Yukawa coupling, $\lambda _t$, and the 
squark soft 
masses.
The one loop Renormalization Group Equation (RGE)
 for $m^2_{H_u}$ is (neglecting 
gaugino and the trilinear scalar term
$(H_u \tilde{Q} \tilde{u} ^c)$ contributions ) 
\begin{equation}
\frac{dm_{H_u}^2(t)}{dt} \approx \frac{3 \lambda _t^2}
{8 \pi ^2}
(m_{H_u}^2(t)+m_{\tilde{u} ^c}^2(t) +m_{\tilde{Q}}^2(t)),
\end{equation}
which gives
\begin{equation}
m_{H_u}^2 (t \approx 
\ln (\frac{ m_{\tilde{t}} }{ \Lambda _{mess} })) \approx 
m_{H_u}^2(0)- 
\frac{3 \lambda _t^2}{8 \pi ^2}
( m_{H_u}^2(0)+ m_{\tilde{u} ^c}^2(0) + m
_{\tilde{Q}}^2(0) )
\ln(\frac{\Lambda_{mess}}{m_{\tilde{t}}}).
\label{approx}
\end{equation}
On the right-hand side of equation \ref{approx} 
the 
RG scaling of $m_{\tilde{Q}}^2$ and $m_{\tilde{u} ^c}^2$ 
has been neglected.
Since the logarithm 
$|t|\approx$$\ln(\Lambda_{mess}/m_{\tilde{t}})$ 
is small, it is
naively expected 
that $m^2_{H_u}$ will not be driven
negative enough and will not trigger electroweak symmetry 
breaking. 
However since the squarks are $\approx$ 500 GeV (1 TeV) 
for a
messenger scale $\Lambda_{mess}=$ 50 TeV (100 TeV), 
the radiative
corrections from virtual top squarks are large since 
the squarks are
heavy. A numerical solution of 
the one loop RGE (including gaugino 
and the trilinear scalar term
$(H_u \tilde{Q} \tilde{u} ^c)$ contributions) 
determines $-m^2_{H_u}=$(275 GeV)$^2$  
((550 GeV)$^2$) for
$\Lambda_{mess}=$50 TeV (100 TeV) and setting $\lambda_t=1$. 
Therefore, $m^2_Z/(2|m^2_{H_u}|)\sim$0.06 (0.01), an 
indication of the fine tuning required.

To reduce the fine tuning in the Higgs sector, it
is necessary to reduce $|m^2_{H_u}|$; ideally
so that $m^2_{H_u}\approx-$0.5$m^2_Z$.
The large value of $|m^2_{H_u}|$ at the weak scale 
is a consequence of
the large hierarchy in
the soft scalar masses at the
messenger scale: $m^2_{\tilde{e} _R} <
m_{H_u}^2 \ll m_{\tilde{Q},\tilde{u} ^c}^2$. Models of
sections \ref{toymodel},\ref{NMSSM}, and \ref{complete} 
attempt to reduce the ratio
$m^2_{\tilde{Q}}/m^2_{H_u}$ at the messenger scale 
and hence 
improve the fine tuning in the Higgs sector.

The fine tuning may be
quantified by applying one of the criteria of 
\cite{barbieri1,anderson}. 
The value $O^{\ast}$ of a physical
observable $O$ will depend on the fundamental 
parameters $(\lambda_i)$
of the theory. The fundamental parameters of the 
theory are to be distinguished from the free parameters of the
theory which parameterize the solutions to $O(\lambda_i)=
O^{\ast}$. 
If the value $O^{\ast}$
is unusually sensitive to the
underlying parameters $(\lambda_i)$ of the theory then a small
change in $\lambda_i$ produces a large change in the value of
$O$. The Barbieri-Giudice function
\begin{equation}
c(O,\lambda_i)=\frac{\lambda_i^{\ast}}{O^{\ast}}
\frac{\partial O}{\partial \lambda_i} \big | _{O=O^{\ast}}
\end{equation}
quantifies this sensitivity \cite{barbieri1}. This particular
value of $O$ is fine tuned if the sensitivity to $\lambda_i$
is larger at $O=O^{\ast}$ 
than at other values of $O$ \cite{anderson}.
If there are values of $O$ 
for which the sensitivity to $\lambda_i$
is small, then it is probably sufficient to
use $c(O,\lambda_i)$ as the measure of fine tuning.

To determine $c(m^2_Z,\lambda_i)$, we
performed the following. 
The sparticle spectrum in the
minimal LEGM model is determined by the four parameters
$\Lambda_{mess}$, $\mu^2_3$, $\mu$, and $\tan\beta$.
\footnote{We allow for an arbitrary 
$\mu^2_3$ at $\Lambda_{mess}$.}
The scale $\Lambda_{mess}$ fixes the boundary condition for
the soft scalar masses, and an implicit dependence on
$\tan\beta$ from $\lambda_t$, $\lambda_b$ and $\lambda_{\tau}$ 
arises in RG scaling\footnote{The RG 
scaling of
$\lambda_{t}$ was neglected.}
from $\mu_{RG}=\Lambda_{mess}$ 
to the weak
scale, that is chosen to be 
$\mu^2_{RG}=m^2_t+\frac{1}{2}(\tilde{m}^2_{t}+\tilde{m}^2_{t^c})$. 
The extremization conditions 
of the scalar potential (equations \ref{mssm1} and \ref{mssm2})
together with $m_Z$ and $m_t$ 
leave two free parameters that we choose to be 
$\Lambda_{mess}$ and $\tan\beta$ (see appendix for the expressions 
for these functions). 

A numerical analysis yields the value of 
$c(m^2_Z,\mu^2)$ that is displayed in figure 1
in the 
$(\tan\beta,\Lambda_{mess})$ plane. 
We note that $c(m^2_Z,\mu^2)$ 
is large throughout 
most of the parameter space, except for the region where
$\tan\beta \; \gtap \;$5 and the messenger scale is
 low. A strong
constraint on a lower limit for $\Lambda_{mess}$
is from the right-handed selectron mass. Contours 
$m_{\tilde{e}_R}=$ 75 GeV ($\sim$ the LEP limit from 
the run at
$\sqrt{s} \approx 170$ GeV 
\cite{aleph}) and 85 GeV ($\sim$ the ultimate LEP2 limit 
\cite{cerngroup2})
are also 
plotted. 
The (approximate) limit on the neutralino masses from
the LEP run at $\sqrt{s} \approx 170$ GeV, 
$m_{\chi ^0_1} + m_{\chi ^0_2} = 160$ GeV and the ultimate 
LEP2 limit, 
$m_{\chi ^0_1} + m_{\chi ^0_2} \sim 180$ GeV are also
shown in figures a and c for $sgn(\mu)=-1$ and figures 
b and d for $sgn(\mu)=+1$.
The constraints
from the present and the ultimate LEP2 limits
on the chargino mass are weaker than 
or comparable to those from
the selectron and the neutralino 
masses and are therefore not shown.
If $m_Z$ were much larger, then $c\sim$ 1.
For example,
with $m_Z=$ 275 GeV
(550 GeV) and $\Lambda_{mess}$= 50 (100) TeV, 
$c(m^2_Z;\mu^2)$ varies between 1 and 5  
for $1.4 \; \ltap \; \tan\beta \; \ltap \; 2$, and is 
$\approx 1$ for
$\tan\beta >2$. This suggests that the interpretation 
that a large value for $c(m^2_Z;\mu^2)$ implies that $m_Z$ is 
fine tuned is probably correct.

From figure 1 we conclude 
that in the minimal LEGM model a fine tuning of approximately
$7\%$ in the Higgs potential is required to produce the 
correct value for $m_Z$. Further, for this fine tuning 
the parameters of the model are restricted to 
the region $\tan \beta \; \gtap$ 5 
and $\Lambda_{mess}\approx$ 45 TeV, corresponding to 
$m_{\tilde{e}_R}\approx$ 85 GeV.
We have also checked that adding more complete 
$(5+\bar{5})$'s does not
reduce the fine tuning.



\section{A Toy Model to Reduce Fine Tuning}
\label{toymodel}

\subsection{\it Model}

In this section the particle content 
and couplings in the messenger sector 
that are suffucient to reduce $|m_{H_u}^2|$ 
is discussed. The aim is
to reduce $m_ {\tilde{Q}} ^2/m_{H_u}^2$ at the scale 
$\Lambda _{mess}$.

The idea is to increase the number of messenger leptons
($SU(2)$ doublets) relative
to 
the number of messenger quarks ($SU(3)$ triplets). 
This reduces both $m_{\tilde{Q}}^2/m_{H_u}^2$ and 
$m_{\tilde{Q}}^2/m^2_{\tilde{e}_R}$
at 
the scale $\Lambda _{mess}$ (see equation \ref
{scalarmass}).
This leads to a smaller value of $|m_{H_u}^2|$ in the 
RG scaling
(see equation \ref{approx}) and the scale $\Lambda_{mess}$
can be lowered since $m_{\tilde{e}_R}$ is larger. 
For example, 
with three doublets and one triplet at a scale 
$\Lambda_{mess} = 30$ TeV, 
so that $m_{\tilde{e} _R} \approx 85$ GeV, we 
find $ |m_{H_u}^2(m_{\tilde{Q}})| \approx (100 \hbox{GeV})^2$ for 
$\lambda_t=1$. 
This may be achieved by the following superpotential 
in the messenger sector
\begin{eqnarray}
W & = & \lambda_{q_1} S q_1 \bar{q_1} + 
\lambda_{l_1}S l_1 \bar{l_1} 
+ \lambda_{l_2}Sl_2 \bar{l_2} 
+ \lambda_{l_3}Sl_3 \bar{l_3}
+\frac{1}{3}\lambda_SS^3 \nonumber \\
 & &  + \lambda_{\phi}S \phi _- \phi _+ 
 + \frac{1}{3}\lambda_NN ^3 
+ \lambda_{q_2} N q_2 \bar{q_2} + 
\lambda_{q_3} N q_3 \bar{q_3},
\label{3doublets}
\end{eqnarray}
where $N$ is a gauge singlet.
The two pairs of triplets $q_2, \bar{q} _2$ and 
$q_3, \bar{q} _3$
are required at low 
energies to maintain
gauge coupling unification.
In this model the additional leptons $l_2,\bar{l}_2$ and
$l_3,\bar{l}_3$
 couple to the singlet $S$, whereas the additional
quarks couple to a different singlet $N$ that does not
couple to the messenger fields $\phi_+$, $\phi_-$.
This can be enforced by 
discrete symmetries (we discuss such a model in
section \ref{complete}).
Further, we assume the discrete charges also forbid
any couplings between $N$ and $S$ at the renormalizable
 level (this is true of the model
in section \ref{complete}) so that SUSY 
breaking is communicated first to $S$ and to $N$ 
only at a higher loop level. 

\subsection{\it Mass Spectrum}

Before quantifying the fine tuning in this model, 
the mass spectrum of the 
additional states is briefly discussed. 
While these fields form complete representations of
$SU(5)$, they are not degenerate in mass.
The vev and $F$-component of 
the singlet $S$
gives a mass $\Lambda_{mess}$
to the messenger lepton multiplets if the $F$-term
splitting between the scalars is neglected. 
As the
squarks in $q_i+\bar{q_i}$ ($i$=2,3) do not couple to 
$S$, they acquire a soft scalar mass
from the same two loop diagrams that are
responsible for the masses of the
MSSM squarks, yielding
$m_{\tilde{q}}\approx
\alpha_3(\Lambda_{mess}) \; \Lambda_{SUSY}/(\sqrt{6} \pi)$.
The fermions in $q+\bar{q}$ also acquire mass at this
scale since, if either $\lambda _{q_2}$ or 
$\lambda _{q_3} \sim \;O(1)$, a 
negative value for $m^2_N$ (the soft scalar mass squared
of $N$) is
 generated from the $\lambda _q N q \bar{q}$ coupling 
at one loop
and thus a
vev for $N \sim$ $m_{\tilde{q}}$ is generated. The 
result
is $m_l/m_q\approx\sqrt{6}\pi/\alpha_3
(\Lambda_{mess})(\Lambda_{mess}/\Lambda_{SUSY})
\approx85$.

The mass splitting in the extra fields introduces a
threshold correction to $\sin^2\theta_W$ if it is 
assumed that
the gauge couplings unify at some high scale
$M_{GUT}\approx$10$^{16}$ GeV. 
We
estimate that the splitting shifts the prediction for
 $\sin^2\theta_W$ by an amount
$\approx-$7$\times$ 10$^{-4} \ln(m_l/m_q) n$, 
where $n$ is
the number of split $(5+\bar{5})$.\footnote{The
complete $(5+\bar{5})$, {\it i.e.}, $l_1, 
\bar{l} _1 \; \hbox{and} \;
q_1, \bar{q} _1$, that couples to $S$
is also split because $\lambda_l\neq\lambda_q$ at 
the messenger scale
due to RG
scaling from $M_{GUT}$ to $\Lambda_{mess}$.
This splitting
is small and neglected.} In this case
$n=$2 and $m_l/m_q \sim$ 85, 
so $\delta$$\sin^2\theta_W\sim-6\times 10^{-3}$.
If
$\alpha_3(M_Z)$ and $\alpha_{em}(M_Z)$ are used
as input, then using the two loop RG equations
$\sin^2\theta_W(\overline{MS})
=0.233 \; \pm \;O(10^{-3})$
is predicted in a minimal SUSY-GUT \cite{langacker}.
The error is a combination of 
weak scale SUSY and GUT 
threshold corrections\cite{langacker}. 
The central value of the
theoretical prediction is a few percent higher
than the measured value of
$\sin^2\theta_W(\overline{MS})=0.231 \pm 0.0003$\cite{pdg}.
The split extra fields shift the prediction 
of $\sin^2\theta_W$ to $\sim 0.227 \pm\;O(10^{-3})$ which is a few 
percent lower than the experimental value. In sections 
\ref{GUT},\ref{complete} we show that this spectrum is derivable from 
a $SU(5)\times SU(5)$ GUT in which 
the GUT threshold 
corrections to $\sin^2\theta_W$ 
could be $\sim O(10^{-3})-O(10^{-2})$ \cite{barr}. It is possible that 
the combination of these GUT threshold corrections and the split
extra field threshold corrections make the prediction of 
$\sin^2\theta_W$ more 
consistent with the observed value.

\subsection{\it Fine Tuning}
\label{Fine tuning}
To quantify the fine tuning in these class of models the 
analysis of section 3 is applied. In our RG analysis the 
RG scaling of $\lambda_t$, the effect of the extra 
vector-like 
triplets on the RG scaling of the gauge couplings, and weak 
scale SUSY threshold corrections were neglected. We have 
checked 
{\it a postiori} that this approximation is consistent. 
As in section 
\ref{finetune}, the two free parameters are 
chosen to be $\Lambda_{mess}$ and $\tan\beta$.
Contours of constant $c(m^2_Z,\mu^2)$ are 
presented in figure 2. 
We show contours of 
$m_{\chi^0_1} + m_{\chi^0_2}= 160$ GeV, 
and
$m_{\tilde{e}_R}= 75$ GeV in figure
2 a for $sgn(\mu)=-1$ and 2b for $sgn(\mu)=+1$.
These are roughly the
present limits from LEP (including the run at
$\sqrt{s} \approx 170$ GeV \cite{aleph}).
The (approximate) 
ultimate LEP2 reaches \cite
{cerngroup2} 
$m_{\chi^0_1} + m_{\chi^0_2} = 180$ GeV,
and
$m_{\tilde{e}_R}= 85$ GeV are shown in figure 2c for
$sgn(\mu)=-1$ and figure 2d for $sgn(\mu)=+1$.
Since $\mu^2 (\approx$ (100 GeV)$^2)$ is much smaller 
in these models than in the minimal LEGM model, 
the neutralinos ($\chi^0_1 \; \hbox{and} \;
\chi^0_2$) are lighter 
so that the neutralino masses
provide a stronger constraint on 
$\Lambda_{mess}$ than does the slepton mass 
limit. The chargino constraints are comparable to the 
neutralino constraints and are thus not shown. It is 
clear that there 
are areas of parameter space in which 
the fine tuning is improved to $\sim$ 40$\%$
(see figure 2). 



While this model improves the fine tuning required 
of the $\mu$ parameter, it would be unsatisfactory 
if further fine tunings were required in other sectors 
of the model, for example, the sensitivity of
$m^2_Z$ to $\mu^2_3$, $\Lambda_{mess}$ and 
$\lambda_t$ and the 
sensitivity of $m_t$ to $\mu^2$, $\mu^2_3$,
$\Lambda_{mess}$ and $\lambda_t$. 
We have checked that all these are less than 
or comparable to $c(m^2_Z;\mu^2)$. We
now discuss the other fine tunings in detail.

For large $\tan\beta$, the sensitivity of $m^2_Z$
to $\mu _3 ^2$,
$c(m^2_Z;\mu^2_3)$ $\propto$ $1/\tan^2\beta$, and 
is therefore smaller than $c(m^2_Z;\mu^2)$.
Our numerical analysis shows that 
$c(m^2_Z;\mu^2_3) \; \ltap \; c(m^2_Z;\mu^2)$
for all $\tan \beta$.
 
In the one loop 
approximation $m^2_{H_u}$ and $m^2_{H_d}$
at the weak scale are proportional 
to  $\Lambda^2 _{mess}$ since 
all the soft masses 
scale with $\Lambda _{mess}$ 
and there is only a weak logarithmic dependence
on $\Lambda _{mess}$ through the gauge couplings. 
We have checked 
numerically 
that $(\Lambda _{mess} ^2/m^2_{H_u}) 
(\partial m^2_{H_u}/ \partial \Lambda _{mess} ^2) \sim 1$. 
Then, $c(m_Z^2;\Lambda _{mess} ^2) \approx
c(m_Z^2;m^2_{H_d}) + c(m_Z^2;m^2_{H_u})$. We find that 
$c(m_Z^2;\Lambda _{mess} ^
2) \approx c(m_Z^2; \mu ^2)+$1 over most of the 
parameter space.

In the one loop approximation, $m^2_{H_u}(t)$ is 
\begin{equation}
m^2_{H_u}(t) \approx m^2_{H_u}(0)
+(m^2_{ \tilde{Q} _3}(0)+m^2_{\tilde{u} ^c_3}(0)
+m^2_{H_u}(0))(e^{-\frac{3 \lambda^2_t}{8 \pi^2}t}-1).
\end{equation} \\
Then, using
$t\approx \ln(\Lambda_{mess}/m_{ \tilde{Q} _3})\approx
\ln(\sqrt{6} \pi/\alpha_3)\approx$ 4.5 and  
$\lambda_t \approx$ 1, 
$c(m^2_Z; \lambda_t)$ is (see appendix)
\begin{equation}
c(m^2_Z; \lambda_t) \approx \frac{4}{m^2_Z}
\frac{\partial m^2_{H_u}(t)}{\partial \lambda ^2_t} 
\approx 50 \frac{m^2_{ \tilde{Q} _3}}{(\hbox{600 GeV})^2}.
\end{equation}
This result measures the sensitivity of $m^2_Z$ to 
the value of $\lambda_t$ at the electroweak scale. 
While this 
sensitivity is large, it does not reflect the fact that 
$\lambda_t(M_{pl})$ is the fundamental parameter of 
the theory, 
rather than $\lambda_t(M_{weak})$.
We find by
 both numerical and 
analytic computations that, for this model 
with three $(5+ \bar{5})$'s 
in addition to the MSSM particle content, 
$\delta \lambda_t(M_{weak}) 
\approx 0.1 \times \delta \lambda_t(M_{pl})$, and 
therefore  
\begin{equation}
c(m^2_Z; \lambda_t(M_{pl})) \approx 5
\frac{m^2_{ \tilde{Q} _3}}
{(\hbox{600 GeV})^2}.
\end{equation} 
For a scale of $\Lambda_{mess}$
= 50 TeV ($m_{\tilde{Q}} \approx$ 600 GeV),
$c(m^2_Z; \lambda_t(M_{pl}))$ is comparable to 
$c(m^2_Z;\mu^2)$ which is
$\approx$ 4 to 5.
At a lower messenger scale, $\Lambda_{mess} \approx$ 35 TeV, 
corresponding to
 squark masses 
of $\approx$ 450 GeV, the sensitivity of $m^2_Z$ to 
$\lambda_t(M_{pl})$ is
$\approx$ 2.8. This is comparable to $c(m^2_Z; \mu^2)$ 
evaluated 
at the same scale.

We now discuss the sensitivity of $m_t$ to the fundamental 
parameters.
Since, $m_t^2  = \frac{1}{2} v^2 \sin ^2 \beta \lambda _t ^2$, 
we get
\begin{equation}
c(m_t; \lambda _i) = \delta _{\lambda _t \lambda _i} +
\frac{1}{2}c(m_Z^2; \lambda _i) 
+ \frac{\cos ^3 \beta}{\sin \beta} \frac{\partial \tan \beta}
{\partial \lambda _i}\lambda_i.
\end{equation}
Numerically
we find that the last term in $c(m_t; \lambda _i)$ 
is small compared to 
$c(m_Z^2; \lambda _i)$
and thus over most of 
parameter space $c(m_t; \lambda _i) \approx
\frac{1}{2}c(m_Z^2; \lambda _i)$. 
As before, the sensitivity of $m_t$ 
to the value of $\lambda _t$ at the GUT/Planck scale 
is much smaller than
the sensitivity to the value of $\lambda _t$ at the weak scale.

\subsection{\it Sparticle Spectrum}

The sparticle spectrum is now briefly discussed to
highlight deviations from the mass relations
predicted in the minimal LEGM model. For example, with three
doublets and one triplet at a scale of
$\Lambda=$ 50 TeV, the soft scalar masses (in GeV) at a
renormalization scale $\mu^2_{RG}=m^2_t
+\frac{1}{2}(m^2_{ \tilde{Q} _3}+m^2_{\tilde{u}^c_3})$
$\approx($630 GeV$)^2$, for $\lambda_t=$ 1, are
shown in table 1.

\begin{table}
\begin{center}
\begin{tabular}{lllll} \hline
$m_{ \tilde{Q} _{1,2}}$ & $m_{\tilde{u}^c_{1,2}}$ &
$m_{\tilde{d}^c_{i}}$ & $m_{\tilde{L}_i,H_d}$ &
$m_{\tilde{e}^c_i}$ \\ \hline
687 & 616 & 612 &319 & 125 \\ \hline
\end{tabular} 
\end{center}  

\begin{center}
\begin{tabular}{ll} \hline
$m_{\tilde{Q}_3}$ & $m_{\tilde{u}^c_3}$ \\ \hline
656 & 546 \\ \hline
\end{tabular} 
\end{center} 
\caption{Soft scalar masses in GeV
for messenger particle content
of three $(l+\bar{l})$'s and one $q+\bar{q}$ and a scale 
$\Lambda _{mess} = 50$ TeV.}
\label{spectrum}
\end{table}

Two observations that are generic to this type of
model are:
(i) By construction, the spread in the soft
scalar masses is less than in the minimal LEGM model.
(ii) The gaugino masses do not satisfy the one-loop
SUSY-GUT relation $M_i/\alpha_i$ = constant. In this case,
for example, $M_3/\alpha_3:M_2/\alpha_2 \approx$ 1$:$3
and $M_3/\alpha_3:M_1/\alpha_1 \approx$ 5$:$11
to one-loop.

We have also found that for $\tan\beta$ $\gtap$ 3, the Next
Lightest Supersymmetric Particle (NLSP) is one of the 
neutralinos, whereas for $\tan\beta$ $\ltap$ 3, the NLSP 
is the right-handed stau. Further, for 
these small values of $\tan\beta$,
the three right-handed sleptons are degenerate within 
$\approx$ 200 MeV. 

\section{NMSSM}
\label{NMSSM}
In section \ref{finetune}, the
$\mu$ term and the SUSY breaking mass $\mu^2_3$ were
put in by hand. There it was found that these parameters 
had to be fine tuned in order to correctly reproduce the 
observed $Z$ mass. 
The extent to which this is 
a ``problem'' may only be evaluated within a specific model
that generates 
both the $\mu$ and $\mu^2_3$ terms. 

For this reason, in this section a possible way to 
generate both the $\mu$ term
and $\mu^2_3$ term in a manner that requires a minimal  
modification to the model of either section
\ref{mess} or section \ref{toymodel} is discussed. 
The easiest way to generate these 
mass terms is to introduce a singlet $N$ and add 
the interaction
 $N H_u H_d$ to the superpotential (the NMSSM)\cite{fayet}.
The vev of the scalar component of $N$ 
generates $\mu$ and the vev of the $F$-component of
$N$ generates $\mu_3^2$. 

We note that for the
``toy model'' solution to the fine tuning problem 
(section \ref{toymodel}),
the introduction of the singlet occurs at no additional cost.
Recall that in that model  
it was necessary to introduce a singlet $N$, 
distinct from $S$, 
such that the vev of $N$ gives mass 
to the
extra light vector-like triplets, $q_i,\bar{q} _i \; 
(i=2,3)$ (see
equation \ref{3doublets}).
Further, discrete 
symmetries (see section \ref{complete})
are imposed to isolate $N$ from SUSY breaking
in the messenger sector.
This last requirement 
is necessary to solve the 
fine tuning problem: if both the 
scalar and $F$-component of $N$ acquired a vev  
at the same scale as $S$, then the extra triplets that 
couple to
$N$ would also act as messenger fields. 
In this case the messenger fields would form 
complete $(5+\bar{5})$'s and the fine tuning problem 
would be 
reintroduced. With $N$ isolated from the messenger sector 
at tree level, a vev 
for $N$ at the electroweak scale is naturally generated,
as discussed in section \ref{toymodel}.

We also comment on the necessity and origin of these extra
triplets.
Recall that in the toy model of section \ref{toymodel}
these triplets were required to maintain the SUSY-GUT
prediction for $\sin^2\theta_W$. Further, we shall also 
see that
they are required in order to generate a large enough
$-m^2_N$ (the soft scalar mass squared of the singlet $N$). 
Finally, in the GUT model
of section \ref{complete}, the lightness of these triplets
(as compared to the missing doublets) is the consequence
of a doublet-triplet splitting mechanism.

The superpotential in the electroweak symmetry 
breaking sector is  
\begin{equation}
W = \frac{\lambda _N}{3} N^3 
+ \lambda _q N q \bar{q} - \lambda _H N H_u H_d,
\label{WNMSSM}
\end{equation}
which is similar to an  
NMSSM except for the coupling of $N$ to the triplets. 
The superpotential in the messenger sector is 
given by equation \ref{3doublets}.

The scalar potential is
\footnote{In models of gauge mediated 
SUSY breaking, $A_H$=0 at
tree level and a non-zero value
of $A_H$ is generated at one loop. 
The trilinear scalar term $A_N N^3$ is generated at 
two loops
and is neglected.} 
\begin{eqnarray}
V &=& \sum_{i} | F_i | ^2 + m_N^2 | N | ^2 + 
m_{H_u}^2 | H_u | ^2 + 
m_{H_d}^2 | H_d | ^2 +\hbox{D-terms} \nonumber \\
 & & -(A_H NH_uH_d+h.c.)
+V_{1-loop}.
\label{Vscalar}
\end{eqnarray}
The extremization conditions for 
the vevs of the real components of $N$, $H_u$ 
and $H_d$, denoted by
$v_N$, $v_u$ and $v_d$ respectively (with 
$v = \sqrt{v_u^2 +v_d^2} \approx 250$
GeV), are  
\begin{equation}
v_N (\tilde{m}^2_N + \lambda ^2_H \frac{v^2}{2} + 
\lambda ^2_N v^2_N -
\lambda _H \lambda _N v_u v_d) -
\frac{1}{\sqrt{2}}A_Hv_uv_d= 0,
\label{vn}
\end{equation}
\begin{eqnarray}
\frac{1}{2} m_Z^2 & = & 
\frac{ \tilde{\mu} _1 ^2 - \tilde{\mu} _2 ^2 \tan ^2 \beta }
{\tan ^2 \beta - 1 }, 
\label{NMSSM1}
\\
\sin 2 \beta & = &  2 \frac{\mu ^2 _3}
{\tilde{\mu} _2 ^2 + \tilde{\mu} _1 ^2},
\label{NMSSM2}
\end{eqnarray}
with
\begin{eqnarray}
\mu ^2 & = & \frac{1}{2} \lambda _H ^2 v_N^2 ,    \\
\mu _3 ^2 & = & -\frac{1}{2} \lambda _H ^ 2 v_u v_d 
+ \frac{1}{2} 
\lambda _H \lambda _N v_N^2+A_H\frac{1}{\sqrt{2}}v_N, 
\label{Bmu}
\\
\tilde{m}^2_i & = & m^2_i
+2\frac{\partial V_{1-loop}}{\partial v^2_i}
\hbox{}; \; \; i=(u,d,N).
\end{eqnarray}

We now comment on the expected size of the Yukawa couplings 
$\lambda_q$, $\lambda_N$ and $\lambda_H$.
We must use the RGE's to evolve these couplings from their 
values at
$M_{GUT}$ or $M_{pl}$ to the weak scale. The quarks and the
Higgs doublets receive 
wavefunction renormalization from $SU(3)$ and 
$SU(2)$ gauge interactions respectively, whereas the 
singlet $N$
does not receive any
 wavefunction renormalization from gauge interactions at 
one loop.
So, the couplings at the weak scale are in the order:
$\lambda _q \sim O(1) > \lambda _H > \lambda _N$ if 
they all 
are $O(1)$
at the GUT/Planck scale. 

We remark that 
without the $N q \bar{q}$ coupling, it is difficult to 
drive a vev for
$N$ as we now show below.
The one loop RGE for $m_N^2$ is
\begin{equation}
\frac{dm_N^2}{dt} 
\approx \frac{6 \lambda _N^2}{8 \pi ^2} m_N^2(t) +
\frac{2 \lambda _H^2}{8 \pi ^2} (m_{H_u}^2(t)+m_{H_d}^2(t) 
+m_N^2(t)) +
\frac{3 \lambda _q^2}{8 \pi ^2} (m_{\tilde{q}}^2(t) + 
m_{\tilde{
\bar{q}}}^2(t)).
\end{equation}
Since $N$ is a gauge-singlet, 
$m_N^2 =0 $ at $\Lambda _{mess}$. 
Further, if $\lambda _{q} = 0$, an estimate 
for $m_{N}^2$ at the weak scale is then
\begin{equation}
m_N^2 \approx - \frac{2 \lambda ^2_H}{8 \pi ^2} 
(m_{H_u}^2(0) + 
m_{H_d}^2(0)) \ln \left(
\frac{\Lambda _{mess}}{m_{H_d}} \right),
\label{mn1}
\end{equation}
{\it i.e.}, $\lambda _H$ drives $m_N^2$ negative.
The extremization condition for $v_N$, equation 
\ref{vn}, and 
using equations \ref{NMSSM2} and \ref{Bmu} (neglecting $A_H$)
shows that 
\begin{equation}
m^2_N + \lambda ^2_H \frac{v^2}{2}
\approx \lambda ^2 _H \left(\frac{v^2}{2} - 
\frac{2}{8 \pi ^2}
(m_{H_u}^2(0) +m_{H_d}^2(0)) \ln \left(
\frac{\Lambda _{mess}}{m_{H_d}} \right) \right)
\end{equation}
has to be negative for $N$ to acquire a vev. This implies 
that 
$m_{H_u}^2$ and $m_{H_d}^2$ at $\Lambda _{mess}$ 
have to be greater
than $\sim (350 \; \hbox{GeV})^2$ which implies that 
a fine tuning of a few percent is required 
in the electroweak symmetry
breaking sector.
With $\lambda _q \sim O(1)$, however,
there is an additional negative contribution to 
$m_N^2$ given approximately by
\begin{equation}
- \frac{3 \lambda ^2_q}{8 \pi ^2} (m_{q}^2(0) +
m_{\tilde{\bar{q}}}^2(0)) \ln \left( 
\frac{ \Lambda _{mess} }{m_{
\tilde{q} } } \right).
\end {equation}
This contribution dominates the one in equation \ref{mn1}
since
$\lambda _q > \lambda _H$ and the squarks $\tilde{q}$, 
$\tilde{\bar{q}}$ have soft 
masses larger than the
Higgs. 
Thus, with $\lambda _q \neq 0$, 
$m^2_N + \lambda ^2_H v^2/2$ is naturally negative.

Fixing $m_Z$ and $m_t$,
we have the following 
parameters : $\Lambda _{mess}$, 
 $\lambda_q$, $\lambda_{H}$,
 $\lambda_N$, $\tan\beta$,
and $v_N$.
Three of the parameters are fixed by the 
three extremization conditions, leaving three 
free parameters that for convienence are
chosen
to be $\Lambda_{mess}$, $\tan\beta\geq$0,
and $\lambda_H$. The signs of the vevs are fixed to be 
positive by requiring a stable vacuum and no  
spontaneous CP violation.
The three extremization equations determine the following
relations
\begin{eqnarray}
\lambda_N &=&\frac{2}{\lambda_H v^2_N}(\mu^2_3
+\frac{1}{4}\lambda^2_H 
\sin2\beta v^2-\frac{1}{\sqrt{2}}A_Hv_N) ,\\ 
v_N & = & \sqrt {2} \frac{\mu}{\lambda _H}, \\
\tilde{m}^2_N &=& \lambda_N \lambda_H \frac{1}{2}\sin2\beta v^2
-\lambda^2_N v^2_N-\frac{1}{2}\lambda^2_H v^2
+\frac{1}{2 \sqrt{2}}A_H\sin2\beta \frac{v^2}{v_N},
\label{sol}
\end{eqnarray}
where
\begin{eqnarray}
\mu ^2 & = & - \frac{1}{2} m^2_Z 
+ \frac{ \tilde{m}_{H_u}^2 \tan ^2 \beta - \tilde{m}^2_{H_d} }
{ 1 - \tan ^2 \beta } ,\\
2 \mu ^2_3 & = &\sin 2 \beta (2 \mu ^2 + 
\tilde{m}_{H_u}^2 + \tilde{m}^2_{H_d}).
\end{eqnarray}
The superpotential term $NH_uH_d$ couples the RGE's for 
$m^2_{H_u}$, $m^2_{H_d}$ and $m^2_N$. Thus the values of 
these masses
at the electroweak scale are, in general, complicated 
functions of 
the Yukawa parameters $\lambda_t$, $\lambda_H$, 
$\lambda_N$ and 
$\lambda_q$. 
In our case, two of these Yukawa parameters 
($\lambda _q$ and 
$\lambda _N$) are determined by the
extremization equations and a closed form expression for 
the derived 
quantities cannot be found.
To simplify the analysis,
we neglect the dependence of
$m^2_{H_u}$ and $m^2_{H_d}$ on $\lambda_H$ induced in
RG scaling from $\Lambda_{mess}$ to
the weak scale. Then
$m_{H_u}^2$ and $m_{H_d}^2$ depend only on $\Lambda _{mess}$ 
and $\tan \beta$
and thus closed form solutions for
$\lambda_N$, $v_N$ and
$\tilde{m}^2_N$ can be obtained using the above equations.
Once $\tilde{m}_N^2$ at the weak scale is obtained, 
the value of $\lambda _q$ is obtained by using an 
approximate analytic solution. 
An exact numerical solution of the
RGE's then shows that the above approximation is 
consistent.

\subsection{\it Fine Tuning and Phenomenology}

The fine tuning functions we consider below are
$c(O;\lambda_H)$, $c(O;\lambda_N)$, $c(O;\lambda_t)$,
$c(O;\lambda_q)$ and $c(O;\Lambda_{mess})$ where
$O$ is either $m^2_Z$ or $m_t$.
The expressions for the fine tuning functions and other 
details are
given in the appendix. In our RG analysis the approximations
 discussed in 
subsection \ref{Fine tuning} and above were used and 
found to 
be consistent.
Fine tuning contours of 
$c(m^2_Z;\lambda_H)$ are displayed in 
figures 3 a and 3 b for
$\lambda_H=0.1$ and figures 3 c and 3 d 
for $\lambda_H=0.5$.
We have found by numerical computations that
the other fine tuning functions are either smaller or
comparable to $c(m^2_Z; \lambda_H)$. \footnote{
In computing
these functions
the weak scale value of the couplings $\lambda_N$
and $\lambda_H$ has been used.
But since $\lambda_N$
and
$\lambda_H$ do not have a fixed point behavior, we have found
that
$\lambda _{H}(M_{GUT})/\lambda _{H}(m_Z) \;
\partial \lambda _{H}(m_Z)/\partial \lambda _{H}(M_{GUT})
 \sim 1$ so that, for example,
$c( m^2_Z;\lambda_H(M_{GUT}) )\approx
c( m^2_Z;\lambda_H(m_Z) )$.}



We now discuss the existing phenomenological constraints 
on our model
and also the ultimate
constraints if LEP2 does not discover SUSY/light Higgs($h$).
These are shown in figures 3 a,3 c and figures 3 b, 3 d 
respectively.
We consider the processes $e^+ e^-\rightarrow Z h$, 
$e^+ e^-$$\rightarrow (h + pseudoscalar)$,
$e^+ e^-$$\rightarrow$ $\chi^+\chi^-$,
$e^+ e^-$$\rightarrow$ $\chi^0_1\chi^0_2$, and 
$e^+ e^-$$\rightarrow$ $\tilde{e}_R \tilde{e}^{*}_R$ 
observable at LEP.
Since this model also has a light pseudoscalar, 
we also consider upsilon decays
$\Upsilon$$\rightarrow (\gamma +  pseudoscalar)$.
We find that the 
model is phenomenologically viable and requires 
a $\sim$ 20$\%$ tuning even if  
no new particles are discovered at LEP2.

We begin with the constraints on the
scalar and pseudoscalar spectra of this model.
There are three neutral scalars, two neutral 
pseudoscalars and one complex charged scalar.
We first consider the mass spectrum of the pseudoscalars.
At the boundary scale $\Lambda_{mess}$, SUSY is 
softly broken 
in the visible sector only by the soft scalar masses and 
the gaugino 
masses. Further, the superpotential of equation 
\ref{WNMSSM} has
an $R$-symmetry. Therefore,
at the tree level, {\it i.e.,}
with $A_H=$0, the scalar potential
of the visible sector (equation \ref{Vscalar}) has  
a global symmetry.
This symmetry is spontaneously broken
by the vevs of $N^R$, $H^R_u$, and $H^R_d$ (the
superscript $R$ denotes the real component of fields), 
so that one 
physical
pseudoscalar
is massless at tree level. 
It is 
\begin{equation}
a=\frac{1}{ \sqrt{ v^2_N + v^2 \sin ^2 2 \beta } } \left(
v_N N^I + v \sin 2 \beta \cos \beta H_u^I + v \sin 2 \beta
\sin \beta H_d^I \right),
\label{Raxion}
\end{equation}
where the superscripts $I$ denote the imaginary components 
of the 
fields.
The second pseudoscalar, 
\begin{equation}
A\sim- \frac{2}{v_N} N^I + \frac{H_u^I}{v \sin \beta} + 
\frac{H_d^I}{v
\cos \beta},
\end{equation}
acquires a mass
\begin{equation}
m^2_A=\frac{1}{2}\lambda_H\lambda_Nv^2_N(\tan\beta+\cot\beta)
+\lambda_H\lambda_Nv^2\sin2\beta
\label{mA}
\end{equation}
 through the $| F_N | ^2$ term in the scalar potential.

The pseudoscalar $a$ acquires a mass once an 
$A_H$-term is generated, at one loop, through interactions 
with the gauginos.
Including only
the wino contribution in
the one loop RGE, $A_H$ is given by
\begin{eqnarray}
A_H & \approx &
6 \frac{\alpha _2 (\Lambda _{mess})}{4 \pi} M_2 \lambda _H \ln 
\left( \frac{\Lambda _{mess}}{M_2} \right), \nonumber \\
 & \approx & 20 \; \lambda _H  
\left( \frac{M_2}{280 \hbox{GeV}} \right) \hbox{GeV},
\label{A}
\end{eqnarray}
where $M_2$ is the wino mass at the weak scale.
Neglecting the mass mixing between the two pseudoscalars,
the mass of the pseudo-Nambu-Goldstone boson 
is computed to be 
\begin{eqnarray} 
m^2_a & = & \frac{9}{\sqrt{2}} A v_N v_u v_d
/ (v_N^2 + v^2 \sin ^2 2 \beta) \nonumber \\
 & \approx &  (40)^2 \left( \frac{\lambda _H}{0.1} \right)
\frac{M_2}{280 \hbox{GeV}}
\sin 2 \beta 
\left( \frac{ \frac{ {\textstyle v_N} }
{ {\textstyle 250} \hbox{GeV} } }
{\sin^2 2\beta+ \left ( \frac{ {\textstyle v_N} }
{ {\textstyle 250} \hbox{GeV} } \right)^2}\right )
(\hbox{GeV})^2.
\label{ma}
\end{eqnarray}
If the mass of $a$ is less than 7.2 GeV, it could 
be detected in the decay  
$\Upsilon \rightarrow a + \gamma$\cite{pdg}.
Comparing 
the ratio of decay width for
$\Upsilon \rightarrow a + \gamma$ to 
$\Upsilon \rightarrow \mu ^- +\mu ^+$ \cite{pdg,wilczek},
the limit
\begin{equation}
\frac{\sin 2 \beta \tan \beta}
{ \sqrt{ ( \frac{ {\textstyle v_N} }
{ {\textstyle 250} \hbox{GeV} } )^2 
+ \sin ^2 2 \beta } } < 0.43
\label{bb}
\end{equation}
is found.

Further constraints on the spectra are obtained from 
collider searches.
The non-detection of $Z \rightarrow$ scalar + $a$ at 
LEP implies that 
the combined mass of the lightest Higgs scalar and $a$ must 
exceed $\sim$ 92 GeV. 
Also, the process $e^{+}e^{-}$
$\rightarrow$$Zh$ may be observable at LEP2.
For $\lambda _H = 0.1$, 
the constraint $m_h + m_a \; \gtap \;
92$ GeV is stronger than $m_h \; \gtap \; 70$
GeV which is 
the limit 
from LEP at
$\sqrt{s} \approx 170$ GeV \cite{aleph}. 
The contour of $m_h + m_a =
92$ GeV is shown in figure 3 a.
In figure 3 b, we show
the contour of $m_h = 92$ GeV ($\sim$ 
the ultimate LEP2 reach \cite
{cerngroup}).
For $\lambda _H = 0.5$, we find that the constraint 
$m_h \; \gtap \; 70$ GeV
is stronger than $m_h + m_a \; \gtap \; 92$ GeV and restricts
 $\tan \beta \; \ltap \; 5$ independent of $\Lambda _{mess}$.
The contour $m_{h} = 92 $ GeV is shown
in figure 3 d. 
We note that the allowed parameter space is not 
significantly constrained.  
We find that 
these limits make 
the constraint of equation \ref{bb} redundant.
The left-right mixing between the two top squarks was 
neglected in 
computing the top squark radiative corrections to the 
Higgs masses.

The pseudo-Nambu-Goldstone boson $a$ might be 
produced along with the lightest scalar $h$ 
at LEP. 
The (tree-level) 
cross section in units of $R=87/s$ nb
is  
\begin{equation}
\sigma (e^+e^- \rightarrow h\;a) \approx 0.15
\frac{s^2}{(s-m_Z^2)^2} \; \lambda ^2 \; 
v\left(1,\frac{m^2_h}{s},\frac{m^2_a}{s}\right)^3,
\end{equation} 
where $g \lambda /\cos\theta_W$ is the 
$Z(a$$\partial$$h-h$$\partial$$a)$ coupling, and \\
$v(x,y,z)=\sqrt{(x-y-z)^2-4yz}$. 
If $h = c_N N^R + c_u H_u^R + c_d H_d^R $, then
\begin{equation}
\lambda =  \sin 2 \beta \frac{\cos \beta
c_u - \sin \beta c_d}
{ \sqrt{ ( \frac{ {\textstyle v_N} }
{ {\textstyle 250} \hbox{GeV} } )^2
+ \sin ^2 2 \beta }  } .
\end{equation}
We have numerically checked the parameter space allowed 
by $m_h \; \gtap \; 70$ GeV 
and $\lambda_H\leq$0.5 and have found the production 
cross section for 
$ha$ to be less than both the current limit set by DELPHI 
\cite{delphi} 
and a 
(possible) exclusion limit of 30 fb \cite{cerngroup}
at $\sqrt{s} \approx $ 192 GeV.
The production cross-section for $hA$ is larger 
than for $ha$ and $A$ is therefore in principle easier 
to detect.
However, for the 
parameter space allowed by $m_h \; \gtap \; 70$ GeV, 
numerical calculations show that 
$m_A \; \gtap \; $ 125 GeV,
so that this channel is not kinematically accessible.

The charged Higgs mass is
\begin{equation}
m^2_{H ^\pm} = m^2_W + m^2_{H_u} + m^2_{H_d} + 2 \mu ^2
\end{equation}
which is greater than about 200 GeV in this model 
since $m_{H_d}^2 \; \gtap \; (200 \hbox{GeV} )^2$
for $\Lambda _{mess} \; \gtap \; 35$ TeV
and as $\mu ^2 \sim - m^2_{H_u}$.

The neutralinos and charginos may be observable at LEP2 at
$\sqrt{s} \approx 192$ GeV
if $m_{\chi^+} \; \ltap \; 95$ GeV and 
$m_{\chi^0_1}+m_{ \chi^0_2} \; \ltap \; 180$ GeV.
These two constraints are comparable, and thus 
only one of these
is displayed in figures 3 b and 3 d, for 
$\lambda _H = 0.1$ and
$\lambda _H = 0.5$
repectively. Also, 
contours of $m_{\chi^0_1}+m_{ \chi^0_2} =$ 160 GeV ($\sim$
the LEP kinematic limit at  $\sqrt{s} \approx 170$ GeV)
are shown in figures 3 a and 3 c.
Contours of 85 GeV ($\sim$ the ultimate LEP2 limit)
and 75 GeV ($\sim$ the LEP 
limit from $\sqrt{s} \approx 170$ GeV)
for the right-handed selectron mass further constrain the 
parameter space. 

The results presented in all the figures are for a 
central value
of $m_t$=175 GeV. We have varied the top quark 
mass by 10 GeV about the central value of 
$m_t$= 175 GeV and
have found that both the fine tuning measures and the 
LEP2 constraints (the Higgs
mass and the neutralino masses) vary by $\approx$ 30 $\%$, 
but the 
qualitative features are unchanged.

We see from figure 3 that there is 
parameter space allowed by the present limits 
in which the tuning is $\approx$
30 $\%$. Even if no new particles are discovered at
LEP2, the tuning required for some region is
$\approx$ 20$\%$. 

It is also interesting to compare the fine tuning 
measures with those found in the minimal LEGM model
 (one messenger $(5+\bar{5})$)
with an extra singlet $N$ to generate the $\mu$ and
$\mu^2_3$ terms.\footnote{We assume that the
model contains some mechanism to generate 
$-m^2_N\sim(100 \hbox{GeV})^2 - (200 \hbox{GeV})^2$;
for example, the singlet is coupled to an extra 
$(5+\bar{5}$).}
In figure 4 the fine tuning 
contours for $c(m^2_Z;\lambda_H)$ are presented for 
$\lambda_H$=0.1. 
Contours of $m_{\tilde{e} _R} = 75$ GeV and 
$m_{\chi^0_1}+m_{ \chi^0_2} =$ 160 GeV are also shown in 
figure 4 a.  
For $\lambda _H = 0.1$, the
constraint $m_h + m_a \; \gtap \; 92$ GeV is stronger than 
the limit $m_h \; \gtap \; 70$ GeV and is shown in the 
figure 4 a.
In figure 4 b, we show the (approximate)
ultimate LEP2 limits, {\it i.e.,}
$m_h = 92$ GeV, $m_{\chi^0_1}+m_{ \chi^0_2} =$ 180 GeV
and  $m_{\tilde{e} _R} = 85$ GeV. 
Of these constraints, the bound on the lightest Higgs mass
(either $m_h + m_a \; \gtap \; 92$ GeV or $m_h \; 
\gtap \; 92$ GeV) provides
a  strong lower limit on the messenger scale.
We see that in the parameter space allowed by
present limits 
the fine tuning is $\ltap \;2 \%$
and if LEP2 does
not discover new particles, the fine tuning will
be $\ltap \;1 \%$.
The coupling $\lambda_H$ is constrained to 
be not significantly larger than 0.1 if the
constraint 
$m_h + m_a \; \gtap \; 92$ GeV (or $m_h \; \gtap$ 92 GeV) 
is imposed and if the
fine tuning is required to be no worse than 1$\%$.
 


\section{Models Derived from a GUT}
\label{GUT}
In this section, we discuss how the toy model 
of section \ref{toymodel} 
could be derived from a GUT model.

In the toy model of section \ref{toymodel}, the singlets 
$N$ and $S$ do not separately couple to complete $SU(5)$ 
representations (see equation \ref{3doublets}). 
If the extra fields introduced to solve the fine tuning 
problem
were originally part of $(5+\bar{5})$ multiplets,
then the missing triplets (missing doublets)
necessarily couple to the singlet $S(N)$. The triplets
 must
be heavy in order to suppress their contribution to the
soft SUSY breaking mass parameters. If we assume
 the only
other mass scale is $M_{GUT}$, they must acquire a mass
at $M_{GUT}$. This is just the usual problem of 
splitting a $(5+\bar{5})$ \cite{splitting}. 
For example, if the 
superpotential in the messenger sector 
contains four $(5+\bar{5})$'s,  
\begin{equation}
W=\lambda_1S\bar{5}_{l1} 5_{l1} + 
\lambda_2S\bar{5}_{l2} 5_{l2} 
+ \lambda_3S\bar{5}_{l3} 5_{l3} + \lambda_4S\bar{5}_{q} 5_{q},
\end{equation}
then the $SU(3)$ triplets in the $(\bar{5}_l+ 5_l)$'s 
and the $SU(2)$ doublet in $(\bar{5}_q + 5_q)$ must be 
heavy at $M_{GUT}$
so that in the low energy theory there are three doublets 
and one triplet 
coupling to $S$.
This problem can be solved 
using the method of Barbieri, Dvali and Strumia 
\cite{barbieri2} 
that solves 
the usual Higgs doublet-triplet splitting problem. 
The mechanism in this model is
attractive since 
it is possible to make either the doublets or triplets 
of a 
quintet heavy at the GUT scale. We next describe 
their model.
 
 The gauge group is $SU(5) \times 
SU(5) ^{\prime}$, with the particle content
$\Sigma (24,1), \\
\Sigma ^{\prime}(1,24),
\Phi (5, \bar{5}) \, \hbox{and} \,
\bar{\Phi} (\bar{5}, 5) $
and the superpotential can be written as 
\begin{eqnarray}
W&=& \bar{\Phi} ^{\beta} _{\alpha ^{\prime}}
(M_{\Phi} \delta ^{\alpha ^{\prime}}
_{\beta ^{\prime}} \delta ^{\alpha}
_{\beta}+ \lambda \Sigma ^{\alpha}_{\beta} \delta 
^{\alpha ^{\prime}}
_{\beta ^{\prime}}
+\lambda ^{\prime} {\Sigma ^{\prime}}^{\alpha ^{\prime}}
_{\beta^{\prime}} \delta ^{\alpha}
_{\beta}) \Phi ^{\beta ^{\prime}}_{\alpha} 
+ \nonumber \\
 & &  + \frac{1}{2} M_{\Sigma} \hbox{Tr} (\Sigma ^2) 
+ \frac{1}{2} 
M_{\Sigma ^{\prime}} 
\hbox{Tr} (\Sigma^{\prime 2}) + \nonumber \\
 & & \frac{1}{3} \lambda _{\Sigma} \hbox{Tr} \Sigma ^3 +
\frac{1}{3} \lambda _{\Sigma ^{\prime}} \hbox{Tr} 
\Sigma ^{\prime3}.
\end{eqnarray}
A supersymmetric minimum of the scalar potential 
satisfies the 
$F$ - flatness
conditions
\begin{eqnarray}
0&=&F_{\bar{\Phi}}=(M_{\Phi} \delta ^{\alpha ^{\prime}}
_{\beta ^{\prime}} \delta ^{\alpha}
_{\beta}+ \lambda \Sigma ^{\alpha}_{\beta} 
\delta ^{\alpha ^{\prime}}
_{\beta ^{\prime}}
+\lambda ^{\prime} \Sigma
 ^{\prime\alpha^{\prime}}_{\beta ^{\prime}} \delta ^{\alpha}
_{\beta} )
\Phi ^{\beta^{\prime}} _{\alpha} ,\nonumber \\
0&=&F_{\Sigma}=\frac{1}{2} M_{\Sigma} 
\Sigma _{\alpha}^{\beta} + 
\frac{1}{2} \left( \lambda \bar{\Phi} 
_{\alpha ^{\prime}} ^{\beta}
\Phi ^{\alpha ^{\prime}} _{\alpha} - \lambda 
\frac{1}{5} \delta _
{\alpha} ^{\beta} \hbox{Tr} 
(\bar{\Phi} \Phi) \right) 
 + \lambda _{\Sigma} ( \Sigma ^2 - \frac{1}{5} \hbox
{Tr} \Sigma ^2 ) ,\nonumber \\
0&=&F_{\Sigma ^{\prime}}=\frac{1}{2} M_{\Sigma ^{\prime}} 
\Sigma^{\prime\beta^{\prime}}  
_{\alpha ^{\prime}} +
\frac{1}{2} \left(  \lambda ^{\prime} 
\bar{\Phi} ^{\alpha} _{\alpha ^{\prime}}
\Phi ^{\beta ^{\prime}} _{\alpha} - \lambda ^{\prime} 
\frac{1}{5} \delta _
{\alpha ^{\prime}} ^{\beta ^{\prime}} \hbox{Tr} 
(\bar{\Phi} \Phi) \right)
+ \lambda _{\Sigma ^{\prime}} ( \Sigma ^{\prime2} 
- \frac{1}{5} \hbox
{Tr} {\Sigma ^{\prime}}^2 ).
\nonumber \\
\end{eqnarray}
With the ansatz 
\footnote{The 
two possible solutions to the $F$-flatness conditions are
$\Sigma = v_{\Sigma} \, \hbox{diag}(2,2,2,-3,-3)$ and 
$\Sigma = v_{\Sigma} \, \hbox{diag}(1,1,1,1,-4)$.}
\begin{equation}
\Sigma = v_{\Sigma} \, \hbox{diag}(2,2,2,-3,-3) \, ,
\Sigma ^{\prime} 
= v_{\Sigma ^{\prime}} \, \hbox{diag}(2,2,2,-3,-3),
\end{equation}
the $F_{\bar{\Phi}}=0$ condition is 
\begin{equation}
\hbox{diag} [M_3, M_3, M_3, M_2, M_2]\cdot
\hbox{diag} [v_3,v_3,v_3,v_2,v_2]=0,
\end{equation}
where $M_3 = M_{\Phi} + 2 \lambda v_{\Sigma} + 
2 \lambda ^{\prime} 
v_{\Sigma ^{\prime}}$
and $M_2 = M_{\Phi} - 3 \lambda v_{\Sigma} 
- 3 \lambda ^{\prime} v_{\Sigma ^{\prime}}$ and the second 
matrix is 
the vev of $\Phi$. 
To satisfy
this condition, there is a discrete choice for the pattern 
of vev of 
$\Phi$ :
i) $v_3 \neq 0 \, \hbox{and} \, M_3=0 $ or
ii) $v_2 \neq 0 \, \hbox{and} \, M_2=0 $. 
Substituting either i) or ii) in the $F_{\Sigma}$ and 
$F_{\Sigma ^{
\prime}}$ conditions then 
determines
 $v_3$ (or $v_2$).
With two sets of fields, $\Phi _1, \bar{\Phi} _1$ 
with $v_3 \neq 0$ 
and $\Phi _2, \bar{\Phi} _2$ with $v_2 \neq 0$ , we have the 
following pattern
of symmetry breaking
\begin{eqnarray}
SU(5) \times SU(5) ^{\prime} 
& \stackrel{v_{\Sigma},v_{\Sigma ^{\prime}}}
{\rightarrow} &
 (SU(3) \times SU(2) \times U(1))
\times (SU(3) \times SU(2) \times U(1)) ^{\prime} \nonumber \\
  & \stackrel{v_3,v_2}
{\rightarrow}& SM \, 
\hbox{(the diagonal subgroup)}.
\end{eqnarray}
If the scales of the two stages of symmetry breaking 
are about equal,
{\it i.e.} $v_{\Sigma} , \, v_{\Sigma ^{\prime}}
, \, \sim v_3 \, , v_2 \, \sim M_{GUT}$, then
the SM gauge couplings unify at the scale $M_{GUT}$.
\footnote{See \cite{barbieri2} and \cite{barr} 
for models which give this structure of vevs
for the $\Phi$ fields without using the adjoints.}

The particular structure of the 
vevs of $\Phi _1$ and $\Phi _2$ can be 
used to 
split representations as follows. 

Consider the Higgs doublet-triplet splitting problem. 
With the particle
content
$5_h (5,1)$,
 $\bar{5}_h (\bar{5},1)$ and
$X (1,5)$, $\bar{X} (1,\bar{5})$
and the superpotential
\begin{equation}
W = 5 _{h \alpha} \bar{X} ^{\alpha ^{\prime}} 
\bar{\Phi}^{\alpha} _
{1\alpha ^{\prime}} 
+ \bar{5}_h ^{\alpha} X _{\alpha ^{\prime}} {\Phi _1}
 ^{\alpha ^{\prime}} _{\alpha} ,
\end{equation}
the $SU(3)$ triplets in $5_h$, $\bar{5}_h$ and $X$, 
$\bar{X}$ acquire a
mass of order $M_{GUT}$ whereas the doublets in $5_h$, 
$\bar{5}_h$ and 
$X$, $\bar{X}$ are massless. We want only one pair of 
doublets in the 
low energy theory
(in addition to the usual matter fields). The doublets 
in $X$, $\bar{X}$
can be made heavy by a bare mass term $M_{GUT} X \bar{X}$. 
Then
the doublets in 
$5_h, \bar{5}_h$ are the standard Higgs doublets.
But if all
terms consistent with symmetries are allowed 
in the superpotential, then allowing
$M_{GUT} \Phi _1 \bar{\Phi} _1$, $M_{GUT} X \bar{X}$, 
$5_h \bar{X} \Phi _1$ 
and 
$\bar{5}_h X \bar{\Phi} _1$ implies that a bare mass 
term for $5_h \bar{5}_h$ is 
allowed. Of course, we can by hand put in 
a $\mu$ term $\mu 5_h \bar{5}_h$ of 
the order of the weak scale as in section \ref{toymodel}. 
However, 
it is theoretically more desirable to relate all electroweak 
mass scales to the original SUSY breaking scale. 
So, we would like to relate the $\mu$ term to the SUSY 
breaking
scale. We showed in section \ref{NMSSM} that the NMSSM is 
phenomenologically viable and ``un-fine tuned''
in these models.

The vev structure of $\Phi _2$, 
$\bar{\Phi} _2$ can be used to 
make the doublets in a $5 + \bar{5}$ heavy. Again, 
we get two pairs of 
light triplets and one of these 
pairs can be given a mass  at the GUT scale.
  
We can use this mechanism of making either doublets or 
triplets in
a $(5+\bar{5})$ heavy to show how the model of section 
\ref{toymodel} is derivable from 
a GUT. 
The model with three messenger doublets and
one triplet is obtained from a GUT with the following 
superpotential
\begin{eqnarray}
W & = & S 5 \bar{5} + S 5_l \bar{5}_l 
+ S X_l \bar{X}_l + \nonumber \\
 & &  5_l \bar{X}_l \bar{\Phi} _1 
+ \bar{5}_l X_l \Phi _1 + \nonumber  \\
 & &  5_q \bar{X}_q \bar{\Phi} _2 
+ \bar{5}_q X_q \Phi _2 + \nonumber \\
 & &  M_{GUT} X_h \bar{X}_h 
+ 5_h \bar{X}_h \bar{\Phi} _1 
 + \bar{5}_h X_h \Phi _1 
+ \mu 5_h \bar{5} _h\nonumber \\
 & &  + N^3 + N5_q \bar{5} _q 
+ N X_q \bar{X} _q .
\end{eqnarray}
Here, some of the ``extra'' triplets and doublets 
resulting from 
splitting $(5+\bar{5})$'s are massless at the GUT scale. 
For example, the 
``extra'' light doublets are used 
as the additional messenger leptons.
After inserting the vevs and integrating out the heavy 
states, this corresponds to the 
superpotential in equation \ref{3doublets} with the 
transcription:
\begin{eqnarray}
5,\bar{5} & \rightarrow & q_1,\bar{q} _1 
+ l_1,\bar{l} _1 \nonumber \\
5_l, \bar{5}_l 
& \rightarrow & l_2,\bar{l} _2 \nonumber \\
X_l, \bar{X}_l & 
\rightarrow & l_3,\bar{l} _3 \nonumber \\
5_q, \bar{5} _q & 
\rightarrow & q_2, \bar{q} _2 \nonumber \\
X_q, \bar{X} _q & 
\rightarrow & q_3, \bar{q} _3 .
\end{eqnarray}                                                                  
                                              
We conclude this section with a remark about light 
singlets in SUSY-GUT's with low energy 
gauge mediated SUSY breaking.\footnote{The authors thank
H. Murayama for bringing this to 
their attention.}
In a SUSY GUT with a singlet $N$
coupled to the Higgs multiplets,
there is a potential problem of destabilising
the $m_{weak} / M_{GUT}$ hierarchy, if the singlet
is light and if the Higgs triplets have a
SUSY invariant mass of $O(M_{GUT})$ \cite{srednicki}.
In the LEGM models,
a
B-type mass for
the Higgs triplets and doublets is generated at one loop
with gauginos and
Higgsinos in the loop, and with
SUSY breaking coming from the gaugino mass.
Since SUSY breaking (the gaugino mass and the soft scalar
masses) becomes soft above
the messenger scale,
$\Lambda _{mess} \sim$ 100 TeV,
the B-type mass term generated for the Higgs triplets
 is suppressed, {\it i.e.},
it is $O( (\alpha/4 \pi) M_2
\Lambda _{mess}^2 / M_{GUT})$.
Similarly the soft mass squared
for the Higgs triplets are
$O(m_{weak} ^2 \Lambda _{mess}^2 / M_{GUT}
^2)$. Since the triplets couple to the singlet $N$, 
the soft scalar mass and $B$-term generates at one loop
a 
linear term for the scalar and $F$-component of $N$ 
respectively. These tadpoles are harmless since 
the SUSY breaking masses for the triplets are so 
small.
This is to be contrasted with supergravity theories,
where the $B$-term$\sim O(m_{weak}M_{GUT})$ 
and the soft mass $\sim O(m_{weak})$ for the 
triplet Higgs generate a mass for the Higgs doublet
that is at least $\sim O(\sqrt{m_{weak}M_{GUT}}/(4\pi))$.
 
\section{One complete Model}
\label{complete}
The model is based on the gauge group 
$G_{loc}=SU(5) \times SU(5)'$ and 
the global symmetry group 
$G_{glo}=Z_3 \times Z_3' \times Z_4$. The global symmetry acts 
universally on the three
 generations of the SM. The particle 
content and their 
$G_{loc} \times G_{glo}$ quantum numbers are given
in table 2. The most general renormalizable 
superpotential that is consistent with these symmetries 
is
\begin{equation}
W=W_1+W_2+W_3+W_4+W_5+W_6+W_7,
\end{equation}
where,
\begin{eqnarray}
W_1&=&\frac{1}{2}M_{\Sigma}\hbox{Tr}\Sigma^2+
\frac{1}{3}\lambda_{\Sigma}
\hbox{Tr}\Sigma^3
+\frac{1}{2}M_{\Sigma ^{\prime}} \hbox{Tr} \Sigma ^{\prime 2}
+\frac{1}{3}\lambda _{\Sigma ^{\prime}}
\hbox{Tr}\Sigma ^{\prime 3} \nonumber \\
& & +\Phi _2(M_{\Phi _2}+\lambda_{\Phi _2}\Sigma
+\lambda ^{\prime} _{\Phi _2} \Sigma^{\prime}) \bar{\Phi} _2
\nonumber \\
 & & +\Phi _1(M_{\Phi _1}+\lambda_{\Phi _1}\Sigma+
\lambda ^{\prime} _{\Phi _1} \Sigma ^{\prime}) \bar{\Phi} _1,\\
W_2&=&M_1\bar{X}_lX ,\\
W_3&=&\lambda_1\bar{5}_h\Phi _1X_h
+\bar{\lambda}_15_h\bar{\Phi}_1\bar{X}_h
+\lambda_2\bar{5}_l\Phi _1X_l
+\bar{\lambda}_25_l\bar{\Phi}_1\bar{X}_l, \\
W_4&=&\lambda_3\bar{5}_q\Phi _2X_q
+\bar{\lambda}_35_q\bar{\Phi}_2\bar{X}_q ,\\
W_5&=&\lambda_6S5_l\bar{5}_l+\lambda_7S5_q\bar{5}_q
+\lambda_8S\bar{X}_hX_l+\lambda_9S\bar{X}X_h
+\frac{1}{3}\lambda_{S}S^3 ,\\
W_6&=&-\lambda_H 5_h\bar{5}_hN
+\frac{1}{3}\lambda_{N}N^3+
\bar{\lambda} _q N X \bar{X} \nonumber \\
& &+\lambda_{10}N'\bar{X}X_q+\lambda_{11}N'\bar{X}_qX
+\frac{1}{3}\lambda_{N'}N'^3, \\
W_7&=&\lambda^D_{ij}\bar{5}_i10_j\bar{5}_h
+\lambda^U_{ij}10_i10_j5_h.
\end{eqnarray}
The origin of each of the $W_i$'s appearing in the 
superpotential is easy to understand. In computing the
$F$=0 equations at the GUT scale, the only non-trivial 
contributions come 
from fields appearing in $W_1$, since all other $W_i$s 
are bilinear in fields that do not acquire vevs at the 
GUT scale. The function of $W_1$ is to generate the vevs 
$\Sigma,\Sigma'\sim$ diag $[2,2,2,-3,-3]$, 
$\bar{\Phi}^T_2=\Phi _2\sim$ diag $[0,0,0,1,1]$ 
and $\bar{\Phi}^T_1=\Phi _1\sim$ diag $[1,1,1,0,0]$.
These vevs are necessary to break 
$G_{loc}\rightarrow$$SU(3)_c \times SU(2) \times U(1)_Y$
(this was explained in section \ref{GUT}). 
The role of
$W_3$ and $W_4$ 
is to generate the necessary splitting within the 
many $(5+\bar{5})$'s of $G_{loc}$ that is necessary
to solve the usual doublet-triplet splitting problem, as
well as to solve the fine tuning problem that is discussed 
in sections \ref{finetune},\ref{toymodel} and \ref{NMSSM}. 
The messenger sector 
is given by $W_5$. It will shortly be 
demonstrated that at low energies this sector contains 
three vector-like doublets and one vector-like triplet. The 
couplings in $W_6$ and $W_7$ at low energies contain the 
electroweak symmetry breaking sector of the NMSSM, the 
Yukawa couplings of the SM fields, and the 
two light vector-like triplets necessary to maintain 
the few percent prediction for $\sin^2\theta_W$ as well as
to generate a vev for $N$. 

We now show that the low energy theory of this model is the 
model that is discussed in section \ref{NMSSM}.

Inserting the 
vevs for $\Phi _1$ and $\bar{\Phi} _1$ into $W_3$, the 
following 
mass matrix for the colored triplet chiral 
multiplets is obtained:
\begin{equation}
(\bar{5}_h,\bar{X}_h,\bar{5}_l,\bar{X}_l)
\left(\begin{array}{ccccc}
0& \lambda_1 v_{\Phi _1}& 0 &0 &0 \\
\bar{\lambda}_1 v_{\Phi _1} &0 &0 &0 &0  \\
0& 0& 0& \lambda_2 v_{\Phi _1}& 0 \\
0 &0 &\bar{\lambda}_2 v_{\Phi _1} &0 &M_1  \\
\end{array} \right)
\left(\begin{array}{c}
5_h \\
X_h \\
5_l \\
X_l \\
X \\
\end{array} \right)
\end{equation}
and all other masses are zero.
There are a total of four  
vector-like colored triplet fields that 
are massive at $M_{GUT}$.
These are
the triplet components of 
$(5_h,\bar{X}_h)$, $(\bar{5}_h,X_h)$, 
$(\bar{5}_l,X_l)$ and $(\bar{X}_l,T_H)$,
where
$T_H$ is that linear combination of triplets in $5_l$ and 
$X$ that marries the triplet component of $\bar{X}_l$. 
The 
orthogonal combination to $T_H$, $T_L$, is massless at 
this scale.
The massless triplets at $M_{GUT}$ are $(5_q,\bar{5}_q)$, 
$(X_q,\bar{X}_q)$ and $(\bar{X},T_L)$, for a total of 
three vector-like
triplets. By inspection, the only light triplets that 
couple to $S$
at a renormalizable level 
are $5_q$ and $\bar{5}_q$, which was desirable in 
order to solve 
the fine tuning problem. Further, since $X$ contains a 
component of 
$T_L$, the couplings of the other light triplets to the 
singlets 
$N$ and $N'$ are 
\begin{equation}
W_{eff}=\lambda_{10}N'\bar{X}X_q
+\bar{\lambda}_{11}N'\bar{X}_qT_L
+\lambda _q NT_L\bar{X}     ,
\end{equation}
where $\lambda _q = \bar{\lambda} _q \cos \alpha ^{\prime}$,
$\bar{\lambda}_{11}=\lambda_{11} \cos \alpha^{\prime}$ 
and $\alpha^{\prime}$ is the mixing angle 
between
the triplets in 
$5_l$ and $X$, {\it i.e.},
$T_L = \cos \alpha^{\prime} X - \sin \alpha^{\prime} 5_l$. 
The $\lambda _q N T_L \bar{X}$ coupling is also desirable 
to generate
acceptable $\mu$ and $\mu^2_3$ terms (see section 5). 

In section \ref{toymodel},\ref{NMSSM} it was also 
demonstrated that with a total of three messenger doublets 
the fine tuning required in 
electroweak symmetry breaking could be alleviated.
By inserting the vev for $\Phi _2$ into $W_4$, the doublet 
mass matrix is
given as
\begin{equation}
(\bar{X}_l,\bar{5}_q,\bar{X}_q)
\left(\begin{array}{ccc}
M_1 &0 &0 \\
0 &0 &\lambda_3 v_{\Phi _2} \\
0 &\bar{\lambda}_3 v_{\Phi _2} &0 \\
\end{array} \right)
\left(\begin{array}{c}
X \\
5_q \\
X_q \\
\end{array} \right) 
\end{equation}
and all other masses are zero. At $M_{GUT}$ the heavy 
doublets are $(\bar{X}_l,X)$, $(5_q,\bar{X}_q)$ and 
$(\bar{5}_q,X_q)$, leaving the four vector-like doublets in 
$(5_h,\bar{5}_h)$, $(5_l,\bar{5}_l)$, $(\bar{X},X_l)$ and
$(X_h,\bar{X}_h)$ massless at this scale. Of these four pairs, 
$(5_h,\bar{5}_h)$ are the usual Higgs doublets and 
the other three pairs couple to $S$.

The (renormalizable) superpotential at 
scales below $M_{GUT}$ is then
\begin{eqnarray}
W&=&\lambda _q N \bar{q}_2 q_2
+\frac{1}{3}\lambda_{N}N^3+\lambda_{10} N^{\prime} q_3 \bar{q}_2
\nonumber \\
& & +\lambda_{11}N^{\prime} q_2 \bar{q}_3 - \lambda_H N H_u H_d
+\frac{1}{3}\lambda_{N^{\prime}}{N ^{\prime}} ^3 \nonumber \\
& &+\lambda_6S\bar{l}_1l_1
+\lambda_7S\bar{q}_1 q_1 + \lambda_8S\bar{l}_2l_2 
\nonumber \\
& &+\lambda_9 S \bar{l}_3l_3+\frac{1}{3}\lambda_{S}S^3+W_7,
\end{eqnarray}
where the fields have been relabeled to make, in an obvious 
notation,
their $SU(3) \times SU(2) \times U(1)$ quantum numbers apparent.
 
We conclude this section with comments about both the 
choice of 
$Z_4$ as a discrete symmetry and 
about non-renormalizable operators in our model.

The usual $R$-parity violating operators 
$10_{SM}\bar{5}_{SM}\bar{5}_{SM}$ are
not allowed by the discrete symmetries, even at the 
non-renormalizable level. In fact, $R$-parity
 is a good symmetry of the effective theory 
below $M_{GUT}$. By inspection, the fields that acquire 
vevs 
at $M_{GUT}$ are either
invariant under $Z_4$ or have a $Z_4$
charge of $2$ (for example, $\Phi _1$), so that a 
$Z_2$ symmetry is left
unbroken. In fact, the vevs of 
the other fields $S$, $N$, 
$N'$ and the Higgs doublets do not break 
this $Z_2$ either.
By inspecting the $Z_4$ charges of the
SM fields, we see that the unbroken $Z_2$ 
is
none other than the usual $R$-parity. So at $M_{GUT}$, 
the discrete symmetry $Z_4$ is broken 
to $R_p$.
We also note that the $Z_4$ symmetry 
is suffucient to maintain, to 
all orders in $1/M_{Pl}$ operators, the 
vev structure of $\Phi _1$ and $\Phi _2$, 
{\it i.e.}, to forbid
unwanted couplings between 
$\Phi _1$ and $\Phi _2$ that might destabilize 
the vev structure\cite{barr}. This pattern of
vevs  
was essential to solve the 
doublet-triplet splitting problem. 
It is interesting that both 
$R$-parity 
and requiring a viable solution 
to the doublet-triplet splitting problem can be 
accommodated by the same $Z_4$ symmetry.

The non-SM matter fields ({\it i.e.}, the messenger
$5$'s and $X$'s and the light triplets ) have the opposite 
charge to the 
SM matter fields under the unbroken $Z_2$. Thus, there is 
no mass mixing 
between the SM and the non-SM matter fields. 

Dangerous proton decay operators are forbidden 
in this model by the discrete symmetries. Some
higher dimension operators that lead to 
proton decay are allowed, but are suffuciently 
suppressed. We discuss these below.

Renormalizable operators such as
$ 10_{SM} 10_{SM} 5_q$ and 
$ 10_{SM} \bar{5} _{SM} \bar{5} _q$
are 
forbidden by the $Z_3$ symmetries. This is 
necessary to avoid a large proton decay rate.  
A dimension-6 proton decay operator is
obtained by integrating out the colored triplet scalar
components of $5_q$ or $\bar{5}_q$.
Since the colored scalars in $5_q$ and $\bar{5}_q$ 
have a mass $\sim$$O($50 TeV$)$, the presence of these
operators would have led to an unacceptably large 
proton decay rate.

The operators 
$10_{SM} 10_{SM} 10_{SM} \bar{5}_{SM}/M_{Pl}$
and
$10_{SM} 10_{SM} 10_{SM} \bar{5}_{SM} \\ 
(\Phi \bar{\Phi}/M_{Pl}^2)^n /M_{Pl}$,
which give dimension-5 proton decay operators,
are also forbidden by the two $Z_3$ symmetries.
The allowed non-renormalizable operators
that generate dimension-5 proton decay 
operators 
are suffuciently suppressed. 
The operator
$10_{SM} 10_{SM} 10_{SM} \bar{5}_{SM} 
N'/(M_{Pl})^2$,
for example,
is allowed by the discrete symmetries,
but the proton decay rate is safe since 
$v_{N^{\prime}} \sim $ 1 TeV.

The operators 
$10_i 
\bar{5} _j \bar{\Phi} _1 (\bar{X} \; \hbox{or} \; \bar{X}_q) /M_{Pl}$ 
could, in principle, also lead to a large proton
decay rate. Setting $\bar{\Phi} _1$ to its vev, 
the superpotential couplings,
for example,
$\lambda_{ij}(U^c_i D^c_j \bar{X}(\bar{3})
+Q_i L_j \bar{X}(\bar{3}))$ are generated with 
$\lambda_{ij}$ suppressed only by 
$v_{\Phi _1}/M_{Pl}$.
In this model the colored triplet (scalar) components of
$\bar{X}$ and $\bar{X}_q$ have a mass $m_{\tilde{q}} \sim $
500 GeV, giving a  
potentially large proton decay rate. But, in this model 
these operators are forbidden by the discrete symmetries. 
The operator $10_i
\bar{5} _j \bar{\Phi} _1 \bar{X} S/M_{Pl}^2$ is allowed giving a 
four SM fermion proton decay operator with coefficient 
$\sim(v_{\Phi _1} \; v_S /M^2_{Pl})^2 /m_{\tilde{q}}^2 \sim
10^{-34} \hbox{GeV} ^{-2}$. 
This is smaller than the coefficient
generated by exchange of the heavy gauge bosons of mass 
$M_{GUT}$,
which is $\sim g^2_{GUT} / M^2_{GUT} \sim 1/2 \; 10^{-32} 
\hbox{GeV} ^{-2}$ and so this operator leads to
proton decay at a tolerable rate.

With
our set of discrete symmetries, some of the 
messenger states and the light color triplets 
are stable at the renormalizable level.
Non-renormalizable
operators lead to decay lifetime for 
some of these particles of more than about 100
seconds. This is a problem 
from the viewpoint of cosmology, since these
particles decay after Big-Bang Nucleosynthesis (BBN).
With a non-universal choice of discrete symmetries, it 
might be possible
to make these
particles decay before BBN through either 
small renormalizable 
couplings
to the third generation 
(so that the constraints from 
proton decay and FCNC are avoided)
or non-renormalizable operators. This is, however,
 beyond the 
scope of this paper.

\begin{table}
\begin{center}
\begin{tabular}{||l||l|l|l|l||}\hline
$\Psi$ &$\bar{5}_i$ &$10_i$ &$5_h$ &$\bar{5}_h$ \\ \hline
$G_{loc}$ &$(\bar{5},1)$ &$(10,1)$ &$(5,1)$ 
&$(\bar{5},1)$ \\ \hline
$Z_3$ &$1$ &$a$ &$a$ &$a^2$ \\ \hline
$Z'_3$ &$b$ &$1$ &$1$ &$b^2$ \\ \hline
$Z_4$ &$c$ &$c$ &$c^2$ &$c^2$ \\ \hline
\end{tabular}
\end{center}

\begin{center}
\begin{tabular}{||l||l|l|l|l|l|l||}\hline
$\Psi$ &$\Sigma$ &$\Sigma'$ &$\bar{\Phi} _2$ &$\Phi_2$ 
&$\bar{\Phi} _1$ &$\Phi _1$ \\ \hline
$G_{loc}$ &$(24,1)$ &$(1,24)$ &$(\bar{5},5)$ 
&$(5,\bar{5})$ &$(\bar{5},5)$ &$(5,\bar{5})$ 
\\ \hline
$Z_3$ &$1$ &$1$ &$1$ &$1$ &$1$ &$1$ \\ \hline
$Z'_3$ &$1$ &$1$ &$1$ &$1$ &$1$ &$1$ \\ \hline
$Z_4$ &$1$ &$1$ &$1$ &$1$ &$c^2$ &$c^2$ \\ \hline
\end{tabular}
\end{center}

\begin{center}
\begin{tabular}{||l||l|l|l|l|l|l||}\hline
$\Psi$ &$5_l$ &$\bar{5}_l$ &$X_l$ &$\bar{X}_l$ 
&$5_q$ &$\bar{5}_q$   \\ \hline
$G_{loc}$ &$(5,1)$ &$(\bar{5},1)$ 
&$(1,5)$ &$(1,\bar{5})$ &$(5,1)$ &$(\bar{5},1)$ 
 \\ \hline
$Z_3$ &$a^2$ &$1$ &$1$ &$a$ &$1$ &$a^2$ 
  \\ \hline
$Z'_3$ &$1$ &$1$ &$1$ &$1$ &$b^2$ &$b$  \\ \hline
$Z_4$ &$c^2$ &$c^2$ &$1$ &$1$ &$1$ &$1$  \\ \hline
\end{tabular}
\end{center}

\begin{center}
\begin{tabular}{||l||l|l|l|l|l|l||}\hline
$\Psi$ &$X_q$ &$\bar{X}_q$ &$X_h$ &$\bar{X}_h$ &$X$ 
&$\bar{X}$ \\ \hline
$G_{loc}$ &$(1,5)$ &$(1,\bar{5})$
 &$(1,5)$ &$(1,\bar{5})$ &$(1,5)$ &$(1,\bar{5})$ \\ \hline
$Z_3$ &$a$ &$1$ &$a$ &$a^2$ &$a^2$ &$a$ \\ \hline
$Z'_3$ &$b^2$ &$b$ &$b$ &$1$ &$1$ &$b^2$ \\ \hline
$Z_4$ &$1$ &$1$ &$1$ &$1$ &$1$ &$1$ \\ \hline
\end{tabular}
\end{center}
 
\begin{center}
\begin{tabular}{||l||l|l|l|l|l||}   \hline
$\Psi$ &$S$ &$N$ &$N'$ &$\phi_+$ &$\phi_-$\\ \hline
$Z_3$ &$a$ &$1$ &$a$ &$a$ &$a$\\ \hline
$Z'_3$ &$1$ &$b$ &$b^2$ &$1$ &$1$\\ \hline
$Z_4$ &$1$ &$1$ &$1$ &$1$ &$1$\\ \hline
\end{tabular}
\end{center}
\caption{$SU(5) \times SU(5)' \times Z_3
\times Z'_3 \times Z_4$ quantum 
numbers for the fields of the model 
discussed in section 7. $(a,b,c)$ are the generators
of $Z_3 \times Z'_3 \times Z_4$. The three SM generations are
labeled by the index $i$.}
\end{table}

\section{Conclusions}
\label{conclude}
We have quantified the fine tuning 
required in
models of low energy gauge-mediated SUSY breaking to obtain
the correct $Z$ mass. We showed 
that the minimal model requires a fine tuning 
of order $\sim$
7$\%$  
if LEP2 does not discover a
right-handed slepton. We discussed 
how models with more messenger doublets than triplets
can improve the fine tuning.
In particular, a model
with a messenger field particle content of
three $(l+\bar{l})$'s and only one
$(q+\bar{q})$ was tuned to $\sim 40 \%$.
We found that it was necessary to introduce 
an extra singlet to give mass to some color triplets
(close to the weak scale) which are required to 
maintain gauge coupling 
unification. We also discussed 
how the vev and $F$-component of this singlet 
could be used to generate the $\mu$ and 
$B\mu$ terms. We 
found that
for some region of the parameter space this model
requires $\sim \; 25 \%$ tuning and have shown that
limits from LEP do not constrain the parameter space.
This is in contrast to an NMSSM with one
$(5+\bar{5})$ messenger fields, for which we found that
a fine tuning of $\sim \; 1\%$ is required and that
limits from LEP do significantly constrain the
parameter space.

We further discussed how the model with split messenger 
field representations
could be the
low energy theory of a $SU(5) \times SU(5)$ GUT.
A mechanism similar to
the one used to solve the usual
Higgs doublet-triplet splitting problem was used to split
the messenger field representations.
All operators
consistent with gauge and discrete symmetries
were allowed.
In this model $R$-parity is the unbroken subgroup of one
of
the discrete symmetry groups.
Non-renormalizable operators
involving non-SM fields lead to proton decay, but at a safe 
level.

\section{Acknowledgements}
The authors would like to acknowledge N. Arkani-Hamed, C.D. 
Carone, H. Murayama, 
I. Hinchliffe and
M. Suzuki for many useful discussions.
This work was supported in part by the Director, Office of
Energy Research, Office of High Energy and Nuclear Physics,
Division of
High Energy Physics of the
U.S. Department of Energy under Contract
DE-AC03-76SF00098
and in part by the National Science Foundation under
grant PHY-90-21139.
M.G. would also like
to thank the Natural Sciences and Engineering
Research Council of Canada (NSERC) for their support.

\section{Appendix}

In this section the Barbieri-Giudice parameters for
both the MSSM and NMSSM in a gauge mediated 
SUSY breaking scenario are presented.

In an MSSM with gauge mediated SUSY breaking, the
fundamental parameters of the theory (in the 
visible sector) are: $\Lambda_{mess}$;
$\lambda_t$; 
$\mu$; and $\mu^2_3$.
Once electroweak symmetry breaking occurs, the 
extremization conditions determine both $m^2_Z$ and 
$\tan\beta$ as a function of these parameters. To 
measure the sensitivity of $m^2_Z$ to one of the
fundamental parameters $\lambda_i$, 
we compute the variation
in $m^2_Z$ induced by a small change in one of 
the $\lambda_i$. The 
quantity 
\begin{equation}
\frac{\delta m^2_Z}{m^2_Z}\equiv c(m^2_Z;\lambda_i)
\frac{\delta\lambda_i}{\lambda_i},
\end{equation}
where 
\begin{equation}
c(m^2_Z;\lambda_i)=\frac{\lambda_i}{m^2_Z}
\frac{\partial m^2_Z}{\partial\lambda_i},
\end{equation}
measures this sensitivity \cite{barbieri1}. 
In the case of gauge mediated 
SUSY breaking models, there are four functions 
$c(m^2_Z;\lambda_i)$ to be computed. 
They are:

\begin{equation}
c(m^2_Z;\mu^2)=\frac{2\mu^2}{m^2_Z}\left(1+
\frac{\tan^2\beta+1}{(\tan^2\beta-1)^2}\frac{
4\tan^2\beta(\tilde{\mu}^2_1-\tilde{\mu}^2_2)}
{(\tilde{\mu}^2_1-\tilde{\mu}^2_2)(\tan ^2 \beta + 1)
-m^2_Z(\tan^2\beta-1))
}\right),
\end{equation}

\begin{eqnarray}
c(m^2_Z;\mu^2_3)&=&4\tan^2\beta\frac{\tan^2\beta+1}
{(\tan^2\beta-1)^3}\frac{\tilde{\mu}^2_1-\tilde{\mu}^2_2}
{m^2_Z} \nonumber \\
 & \approx & \frac{4}{\tan ^2 \beta}
\frac{\tilde{\mu}^2_1-\tilde{\mu}^2_2}
{m^2_Z}, \; \; \hbox{for large} \; \tan \beta,
\end{eqnarray}

\begin{eqnarray}
c(m^2_Z;\lambda_t)&=& 2 \frac{\lambda _t ^2}{m_Z^2} 
\frac{\partial 
m_Z^2}{\partial m_{H_u}^2}\frac{\partial m_{H_u}^2}
{\partial \lambda _t^2}
\nonumber \\
 & = &\frac{4}{m^2_Z}
 \lambda^2_t \frac{\tan^2\beta}
{\tan^2\beta-1}\frac{\partial m^2_{H_u}}{\partial 
\lambda^2_t}\left(1+2\frac{\tilde{\mu}^2_1-
\tilde{\mu}^2_2}{\tilde{\mu}^2_1+\tilde{\mu}^2_2}
\frac{\tan^2\beta+1}{(\tan^2\beta-1)^2}\right) \nonumber \\
&\approx&\frac{4}{m^2_Z}\frac{\partial m^2_{H_u}}
{\partial \lambda^2_t}, \; \; \; \hbox{for large} \; 
\tan\beta.
\end{eqnarray}
This measures the sensitivity of $m^2_Z$ to the 
electroweak scale value of $\lambda_t$, 
$\lambda_t(M_{weak})$. The Yukawa coupling 
$\lambda_t(M_{weak})$ is not, however, a
 fundamental parameter of the theory. The 
fundamental parameter is the value of the
coupling at the cutoff $\Lambda^0=M_{GUT}$ 
or $M_{pl}$ of
the theory.
We really 
should be computing the sensitivity of 
$m^2_Z$ to this value of $\lambda_t$. 
The measure of sensitivity is then correctly 
given by
\begin{eqnarray}
c(m^2_Z;\lambda_t(\Lambda^0))&=&
\frac{\lambda_t(\Lambda^0)}{\lambda_t(M_{weak})}
c(m^2_Z;\lambda_t(M_{weak}))
\frac{\partial \lambda_t(M_{weak})}{\partial 
\lambda_t(\Lambda^0)}. \\
\end{eqnarray}
We remark that for the model discussed in the 
text with 
three $l+\bar{l}$ and one $q+\bar{q}$ messenger 
fields,
the numerical value of 
$(\lambda_t(\Lambda^0)/\lambda_t(M_{weak}))
\partial \lambda_t(M_{weak})/ 
\partial \lambda_t(\Lambda^0)$ is typically
$\sim 0.1$ because 
$\lambda_t(M_{weak})$ is attracted to its 
infra-red fixed point. This results in a smaller 
value for $c(m^2_Z;\lambda_t)$ than is obtained 
in the absence of these considerations.

With the assumption that $m_{H_u}^2$ and $m_{H_d}^2$ 
scale with 
$\Lambda _{mess} ^2$, we get 
\begin{eqnarray}
c(m_Z^2; \Lambda _{mess} ^2) & = & c(m_Z^2;m_{H_u}^2) 
+ c(m_Z^2;m_{H_d}^2) 
\nonumber \\
 & = & 1 + 2 \frac{\mu ^2}{m_Z^2} 
-\frac{\tan^2\beta+1}{(\tan^2\beta-1)^2} \times 
\nonumber \\
 & & \frac{
4\tan^2\beta (m^2_{H_u}+m^2_{H_d})
(\tilde{\mu}^2_1-\tilde{\mu}^2_2)/m^2_Z}
{(\tilde{\mu}^2_1-\tilde{\mu}^2_2)
(\tan^2\beta+1)-m^2_Z(\tan^2\beta-1)}.
\end{eqnarray}
The Barbieri-Giudice functions for $m_t$ are 
similarly computed. They are

\begin{equation}
c(m_t;\mu^2_3)=\frac{1}{2}c(m^2_Z;\mu^2_3)
+\frac{1}{1-\tan^2\beta},
\end{equation}
\begin{equation}
c(m_t;\mu^2)=\frac{1}{2}c(m^2_Z;\mu^2)
+2\frac{\mu^2}{\tilde{\mu}^2_1+\tilde{\mu}^2_2}
\frac{1}{\tan^2\beta-1},
\end{equation}
\begin{equation}
c(m_t;\lambda_t)=1+\frac{1}{2}c(m^2_Z;\lambda_t)
+\frac{\lambda_t}{\tan^2\beta-1}
\frac{1}{\tilde{\mu}^2_1+\tilde{\mu}^2_2}
\frac{\partial m^2_{H_u}}{\partial \lambda_t},
\end{equation}
\begin{equation}
c(m_t; \Lambda _{mess} ^2) = \frac{1}{2}c(m_Z^2; 
\Lambda _{mess} ^2) 
- \frac{( \tilde{\mu} _1^2 + \tilde{\mu} _2^2 - 
2 \mu ^2 )}
{(1-\tan ^2 \beta) ( \tilde{\mu} _2^2 +\tilde{\mu} _1^2)}.
\end{equation}

Since 
$m_Z$ and $m_t$ are measured, two
of the four fundamental parameters may be eliminated. 
This leaves two free parameters, which for conveinence 
are chosen to be $\Lambda_{mess}$ and $\tan \beta$.

In a NMSSM with gauge mediated SUSY breaking, the 
scalar 
potential for 
$N, \; H_u$ and $H_d$ at the weak scale is specified by the 
following six
parameters: $\lambda _i = m_N^2, m_{H_u}^2, m_{H_d}^2$, 
the $N H_u H_d$ coupling $\lambda _H$,
the scalar $N H_u H_d$ coupling $A_H$, 
and the $N^3$ coupling, $\lambda _N$.
In minimal gauge mediated SUSY breaking, the
trilinear soft SUSY breaking term $NH_uH_d$ is
zero at tree level and is generated at one loop by
wino and bino exchange. In this case,
 $A_H(\lambda_i)=\lambda_H \tilde{A}(\lambda_i)$.
Since the trilinear 
scalar term
$N^3$ is generated at two loops, it is small 
and is 
neglected.
The extremization conditions which determine 
$m_Z = g^2_Z v^2/4 \; (v= \sqrt{v_u^2 + v_d^2}), 
\tan \beta =
v_u/v_d$ and $v_N$ as a function of these 
parameters are given 
in section
\ref{NMSSM}. Equation \ref{vn}  can be written, 
using $\mu = \lambda _H v_N/\sqrt{2}$ as
\begin{equation}
m_N^2 + 2 \frac{\lambda _N^2}{\lambda _H^2} \mu ^2 
- \lambda _H \lambda _N \frac{1}{2} 
v^2 \sin 2 \beta + \frac{1}{2} \lambda _H^2 v^2 - 
\frac{1}{4\mu}
A_H v^2 \lambda _H \sin 2 \beta = 0.
\label{NMSSM1A}
\end{equation}
Equation \ref{NMSSM1} is
\begin{equation}
\frac{1}{8} g_Z^2 v^2 + \mu ^2 - m_{H_u}^2 
\frac{\tan ^2 \beta}{ 1 
- \tan ^2 \beta} 
+ m_{H_d}^2 \frac{1}{1- \tan ^2 \beta} = 0.
\label{NMSSM2A}
\end{equation}
Substituting $v_N^2$ from equation \ref{vn} in 
equation \ref{Bmu} and 
then using this expression for $\mu _3^2$ in 
equation \ref{NMSSM2} gives
\begin{equation}
(m_{H_u}^2 + m_{H_d}^2 + 2 \mu ^2) \sin 2 \beta 
+ \frac{\lambda _H}{\lambda _N} (m_N^2 
+ \frac{1}{2} \lambda _H^2 v^2) + 
A_H (-\frac{2 \mu}{\lambda _H}
- \frac{1}{4} \frac{v^2 \lambda _H ^2 
\sin 2 \beta}{\mu \lambda _N}) = 0.
\label{NMSSM3A}
\end{equation}
The quantity $c = (\lambda_i/m^2_Z)
(\partial m^2_Z / \partial\lambda_i) $
measures the sensitivity of $m_Z$ to these
parameters. This can be computed by differentiating 
equations \ref{NMSSM1A},
\ref{NMSSM2A} and \ref{NMSSM3A}
with respect to these parameters to obtain, after 
some algebra, 
the following 
set of linear 
equations:
\begin{equation}
(A+A_{A_H}) X^i = B^i+B^i_{A_H},
\end{equation}
where
\begin{eqnarray}
A & = & \left( \begin{array} {ccc}
\frac{1}{2} & 1 & \frac{\mu^2_1 - \mu^2_2}{v^2} 
\frac{2 \tan \beta}
{(1 - \tan ^2 \beta )^2} \\
\frac{\lambda _H^3(\lambda _H -\lambda _N \sin2\beta)}
{g_Z^2 \lambda _N^2} 
& 1 & -\frac{1}{2} \frac{\lambda _H^3}{\lambda _N} 
\frac{1 -\tan ^2 \beta}{(1 +\tan ^2 \beta )^2} \\
\frac{v^2}{g_Z^2 (\mu^2_1+\mu^2_2)}
\frac{\lambda _H^3}{\lambda _N} & 
\frac{\sin2\beta v^2}{\mu^2_1+\mu^2_2} & 
\frac{1 -\tan ^2 \beta}{(1 +\tan ^2 \beta )^2}
\end{array} \right) ,\\
A_{A_H}&=&\frac{A_H}{\mu} \times \\
 & & \left(\begin{array}{ccc}
0&0&0\\
-\frac{\lambda^3_H\sin2\beta}{2 g^2_Z \lambda^2_N} &
\frac{\lambda^3_H\sin2\beta}{16 \lambda^2_N}\frac{v^2}{\mu^2}
& \frac{\tan^2\beta-1}{(1+\tan^2\beta)^2}
\frac{\lambda^3_H}{4 \lambda^2_N} \\
-\frac{\lambda^2_H}{2 g^2_Z\lambda_N}
\frac{v^2\sin2\beta}{\mu^2_1+\mu^2_2}
&\frac{v^2}{\mu^2_1+\mu^2_2}(
\frac{\lambda^2_H\sin2\beta}{16 \lambda_N}\frac{v^2}{\mu^2}
-\frac{1}{2\lambda_H})
&\frac{\tan^2\beta-1}{(1+\tan^2\beta)^2}
\frac{\lambda^2_H}{4\lambda_N}
\frac{v^2}{\mu^2_1+\mu^2_2} 
\end{array} \right) ,\nonumber \\
X^{\lambda_H,\lambda_N} & = & \left( \begin{array} {c}
\frac{1}{v^2} \frac{\partial m^2_Z}{\partial \lambda_i} \\
\frac{1}{v^2} \frac{\partial \mu^2}{\partial \lambda_i} \\
\frac{\partial \tan \beta}{\partial \lambda_i} \\
\end{array} \right) ,\\
X^{m^2_i} & = & \left( \begin{array} {c}
\frac{\partial m^2_Z}{\partial m^2_i} \\
\frac{\partial \mu^2}{\partial m^2_i} \\
v^2\frac{\partial \tan \beta}{\partial m^2_i} \\
\end{array} \right) , (i=u,d,N),
\end{eqnarray}
with $\lambda_i=m^2_N,m^2_{H_u},m^2_{H_d},\lambda_H,
\lambda_N$,
and
\begin{eqnarray}
B^{m^2_N}+B^{m^2_N}_{A_H} & = & \left( \begin{array} {c}
0 \\
- \frac{1}{2} \frac{\lambda _H^2}{\lambda _N^2}\\
-\frac{\lambda _H}{\lambda _N} \frac{v^2}{2(\mu^2_1+
\mu^2_2)}\\
\end{array} \right) ,\\
B^{m^2_{H_u}}+B^{m^2_{H_u}}_{A_H} & = & \left( \begin{array} {c}
\frac{\tan ^2 \beta}{1-\tan^2\beta}\\
0 \\
-v^2 \frac{\sin 2 \beta}{2(\mu^2_1+\mu^2_2)}\\
\end{array} \right) ,\\
B^{m^2_{H_d}}+B^{m^2_{H_d}}_{A_H}  & = & \left( \begin{array} {c}
\frac{1}{\tan^2\beta-1} \\
0 \\
-v^2\frac{\sin 2 \beta}{2(\mu^2_1+\mu^2_2)}\\
\end{array} \right) ,\\
B^{\lambda_H} & = & \left( \begin{array} {c}
0\\
-\frac{\lambda _H^3}{\lambda _N^2}
+\frac{3}{4} \frac{\lambda _H^2\sin 2 \beta}{\lambda _N}  
-\frac{\lambda _H}{\lambda _N^2} \frac{m_N^2}{v^2}\\
-\frac{1}{(\mu^2_1+\mu^2_2)} \left( \frac{1}{2}
\frac{m_N^2}{\lambda _N} + \frac{3}{4} v^2 \frac{
\lambda _H^2}{\lambda _N} \right) \\
\end{array} \right) ,\\
B^{\lambda_H}_{A_H}& = & \frac{A_H}{\mu}
\left( \begin{array} {c}
0\\
\frac{\lambda^2_H\sin2\beta}{2 \lambda^2_N} \\
\frac{3}{8}\frac{\lambda_H}{\lambda_N}
\frac{v^2\sin2\beta}{\mu^2_1+\mu^2_2} \\
\end{array} \right) ,\\
B^{\lambda_N} & = & \left( \begin{array} {c}
0\\
-\frac{1}{4} \frac{\lambda _H^3 \sin 2 \beta}{\lambda _N^2}  
+ \frac{\lambda _H^2}{\lambda _N^3} \frac{m_N^2}{v^2}
+\frac{1}{2} \frac{\lambda _H^4}{\lambda _N^3} \\
\frac{1}{2(\mu^2_1+\mu^2_2)} 
\frac{\lambda _H}{\lambda ^2_N} (m_N^2 + \frac{1}{2} v^2 
\lambda _H^2) 
\end{array} \right),\\
B^{\lambda_N}_{A_H}& = & 
\frac{A_H}{\mu}\left( \begin{array} {c}
0\\
-\frac{\lambda^3_H\sin2\beta}{4 \lambda^3_N} \\
-\frac{\lambda^2_H}{8 \lambda^2_N}\frac{v^2\sin2\beta}
{\mu^2_1+\mu^2_2} \\
\end{array} \right). 
\end{eqnarray}
In deriving
these equations
$A_H(\lambda_i)=\lambda_H \tilde{A}(\lambda_i)$ was assumed 
and $\partial \tilde{A} /\partial \lambda_H$ was neglected.
Inverting these set of equations gives the $c$ functions.
We note
that these expressions for 
the various $c$ functions are valid for
any NMSSM in which the $N^3$ scalar term 
is negligible and the $N H_u H_d$ scalar
term is proportional to $\lambda _H$. 
In general, these 6
parameters might, in turn, depend on some fundamental 
parameters,
$\tilde{\lambda} _i$.
Then, the sensitivity to these fundamental parameters is:
\begin{eqnarray}
\tilde{c_i} & \equiv& \frac{\tilde{\lambda}_i}{m^2_Z}
\frac{\partial m^2_Z}{\partial \tilde{\lambda}_i} 
\nonumber \\
 & = & \frac{\tilde{\lambda}_i}{m^2_Z} \sum_{j} 
\frac{\partial \lambda 
_j}{\partial \tilde{\lambda}_i}
\frac{\partial m^2_Z}{\partial \lambda _j} \nonumber \\
& =& \sum_{j}\frac{\tilde{\lambda}_i}{\lambda_j}
c(m^2_Z;\lambda_j)\frac{\partial \lambda_j}
{\partial \tilde{\lambda}_i} .
\end{eqnarray}
For example, in the NMSSM of section \ref{NMSSM}, 
 the fundamental 
parameters are $\Lambda _{mess}, \lambda _H, \lambda _N, 
\lambda _t$
and $ \lambda _q $ ($A_H$ is a function of $
\lambda _H$ and $\Lambda_{mess}$). 
Fixing $m_Z$ and $m_t$ leaves 3 free parameters, 
which we choose to be
$\Lambda _{mess}, \lambda _H$ and $\tan \beta$.
As explained in that section, the effect of 
$\lambda _H$ in the 
RG scaling of $m_{H_u}^2$ and $m_{H_d}^2$ 
was neglected, whereas the sensitivity of 
$m_N^2$ to 
$\lambda _H$ could be non-negligible.
Thus, we have       
\begin{equation}
\tilde{c}(m_Z^2;\lambda _H) = c(m_Z^2;\lambda _H) + 
c(m_Z^2;m_N^2)
 \frac{\lambda _H}{m_N^2} \frac{\partial m_N^2}
{\partial \lambda _H}.
\end{equation}
We find, in our model, that 
$c(m_Z^2;m_N^2)$ is smaller than $c(m_Z^2;\lambda _H)$ 
by a factor
 of $\sim 2$. Also, using approximate analytic and 
also numerical
 solutions to the RG equation
 for $m_N^2$, we find that $(\lambda _H/m_N^2)
(\partial m_N^2/
\partial \lambda _H)$ is $\ltap \; 0.1$.
Consequently, in the analysis of section \ref{NMSSM}
the additional contribution to $\tilde{c}
(m^2_Z;\lambda_H)$
due to the dependence of $m_N^2$ on $\lambda _H$ was 
neglected. A similar conclusion is true for $\lambda _N$.
Also,
\begin{equation}
\tilde{c}(m_Z^2;\lambda _q) = c(m_Z^2;m_N^2) 
\frac{\lambda _q}{m_N^2} \frac{\partial m_N^2}
{\partial \lambda _q}.
\end{equation}
We find that $(\lambda _q/m_N^2) 
(\partial m_N^2/\partial \lambda _q)$ is $\approx 1$ 
so that $\tilde{c}(m_Z^2;\lambda _q)$ is smaller than
$\tilde{c}(m_Z^2;\lambda _H)$ by a factor of 2.

\newpage

Captions:\\
Figure 1: Contours of $c(m^2_Z; \mu^2)=$(10,
15, 20, 25, 40, 60) for a MSSM with a
messenger particle
content of one $(5+\bar{5})$. In figures $(a)$ and
$(c)$ $sgn(\mu)=-$1 and in
figures $(b)$ and $(d)$ $sgn(\mu)=+$1. The
constraints considered are:
(I) $m_{\tilde{e}_R}=$75 GeV , (II)
$m_{{\tilde{\chi}}^{0}_{1}}+m_{{\tilde{\chi}}^{0}_{2}}=$
160 GeV, (III) $m_{\tilde{e}_R}=$85 GeV, and
(IV)
$m_{{\tilde{\chi}}^{0}_{1}}+m_{{\tilde{\chi}}^{0}_{2}}=$
180 GeV.
A central value of
$m_{top}=$175 GeV is assumed.
\\

Figure 2: Contours of $c(m^2_Z;\mu^2)=$(1,
2, 3, 5, 7, 10) for a MSSM with
a messenger particle
content of three $(l+\bar{l})$'s and one
$(q+\bar{q})$.
In figures $(a)$ and
$(c)$ $sgn(\mu)=-$1 and in
figures $(b)$ and $(d)$ $sgn(\mu)=+$1.
The constraints considered are:
(I) $m_{\tilde{e}_R}=$75 GeV , (II)
$m_{{\tilde{\chi}}^{0}_{1}}+m_{{\tilde{\chi}}^{0}_{2}}=$
160 GeV, (III) $m_{\tilde{e}_R}=$85 GeV, and
(IV)
$m_{{\tilde{\chi}}^{0}_{1}}+m_{{\tilde{\chi}}^{0}_{2}}=$
180 GeV.
A central value of
$m_{top}=$175 GeV is assumed.
\\

Figure 3: Contours
of $c(m^2_Z;\lambda_H)$ for the NMSSM of Section 5 and
a messenger particle content of
 three $(l+\bar{l})$'s and one
$(q+\bar{q})$. In figures $(a)$ and $(b)$,
$c(m^2_Z;\lambda_H)$=(4, 5, 6,
 10, 15) and $\lambda_H=$0.1.
In figures $(c)$ and
$(d)$, $c(m^2_Z;\lambda_H)=$(3, 4, 5,
 10, 15, 20) and $\lambda_H$=0.5.
The constraints considered are:
(I) $m_h+m_a=m_Z$, (II) $m_{\tilde{e}_R}=$75 GeV,
(III)
$m_{{\tilde{\chi}}^{0}_{1}}+m_{{\tilde{\chi}}^{0}_{2}}=$
160 GeV,
(IV) $m_h=$ 92 GeV, (V) $m_{\tilde{e}_R}=$85 GeV,
and (VI)
$m_{{\tilde{\chi}}^{0}_{1}}+m_{{\tilde{\chi}}^{0}_{2}}=$
180 GeV. For $\lambda_H=$0.5,
the limit $m_h\gtap$ 70 GeV constrains
$\tan\beta\ltap$ 5 (independent of $\Lambda_{mess}$)
and is thus not shown.
A central value of $m_{top}=$175 GeV
is assumed.
\\

Figure 4: Contours of
$c(m^2_Z;\lambda_H)=$(50, 80, 100, 150, 200)
 for the NMSSM of Section 5 with
$\lambda_H=$0.1 and a messenger
particle content of one $(5+\bar{5})$.
The constraints considered are:
(I) $m_h+m_a=m_Z$,
(II) $m_{\tilde{e}_R}=$75 GeV,
(III)
$m_{{\tilde{\chi}}^{0}_{1}}+m_{{\tilde{\chi}}^{0}_{2}}=$
160 GeV,
(IV) $m_h=$92 GeV,
(V) $m_{\tilde{e}_R}=$85 GeV, and
(VI)
$m_{{\tilde{\chi}}^{0}_{1}}+m_{{\tilde{\chi}}^{0}_{2}}=$
180 GeV.
A central
value of $m_{top}=$175 GeV
is assumed.

\begin{figure}
\leavevmode
\psfig{file=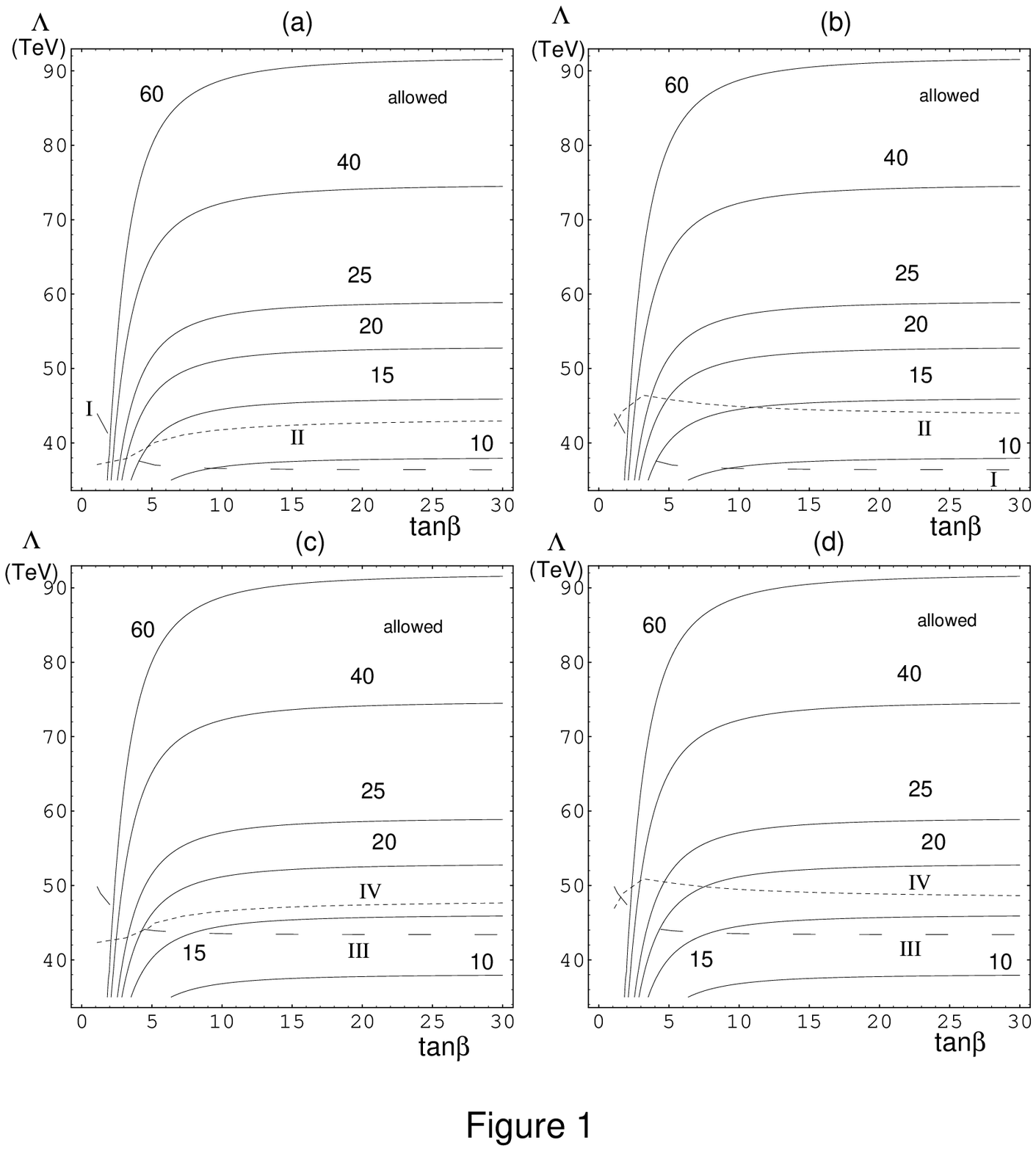}
\protect\label{ftdns1}
\end{figure}

\begin{figure}
\leavevmode
\psfig{file=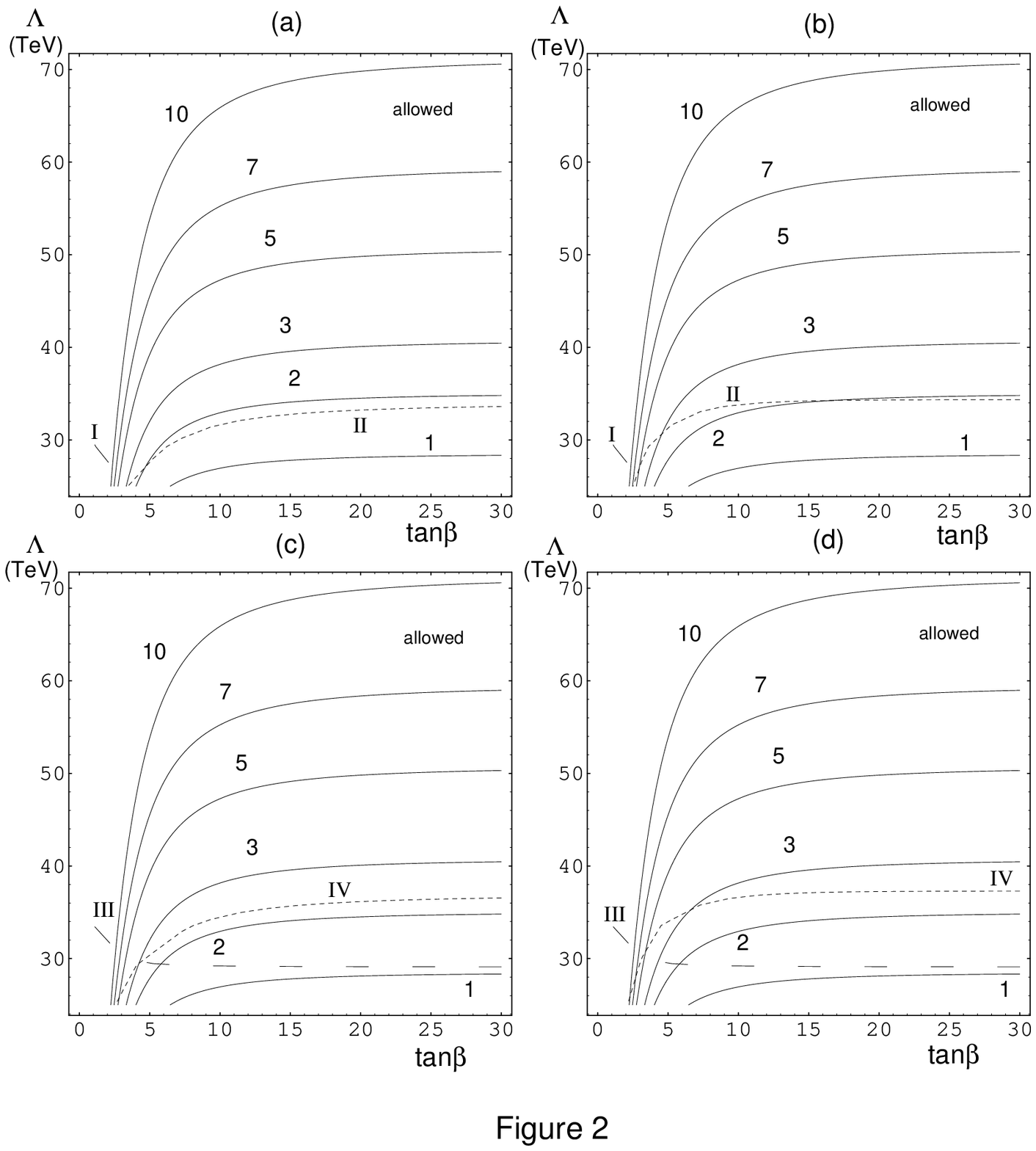}
\protect\label{ft1}
\end{figure}

\begin{figure}
\leavevmode
\psfig{file=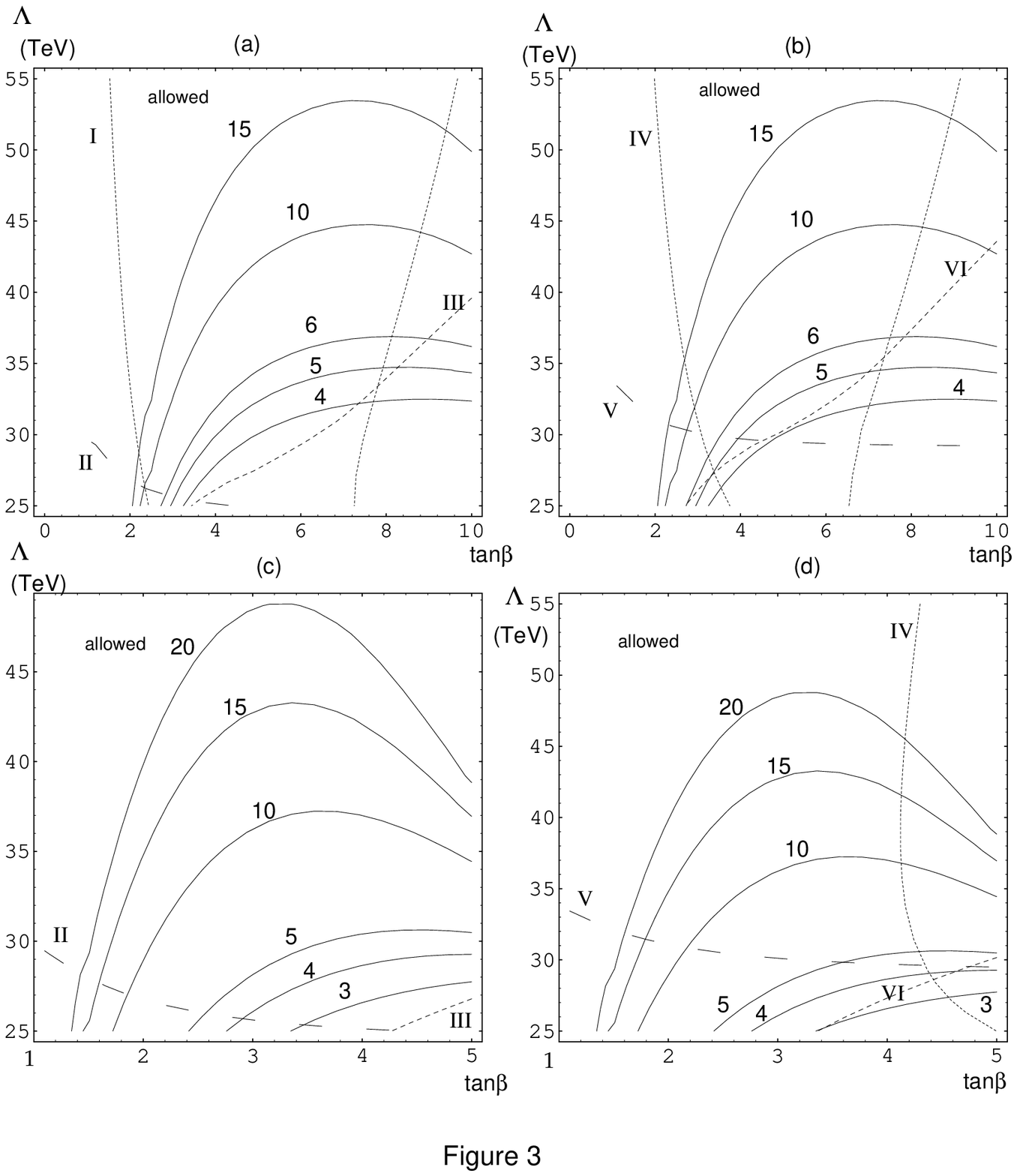}
\protect\label{ftN1}
\end{figure}

\begin{figure}
\leavevmode
\psfig{file=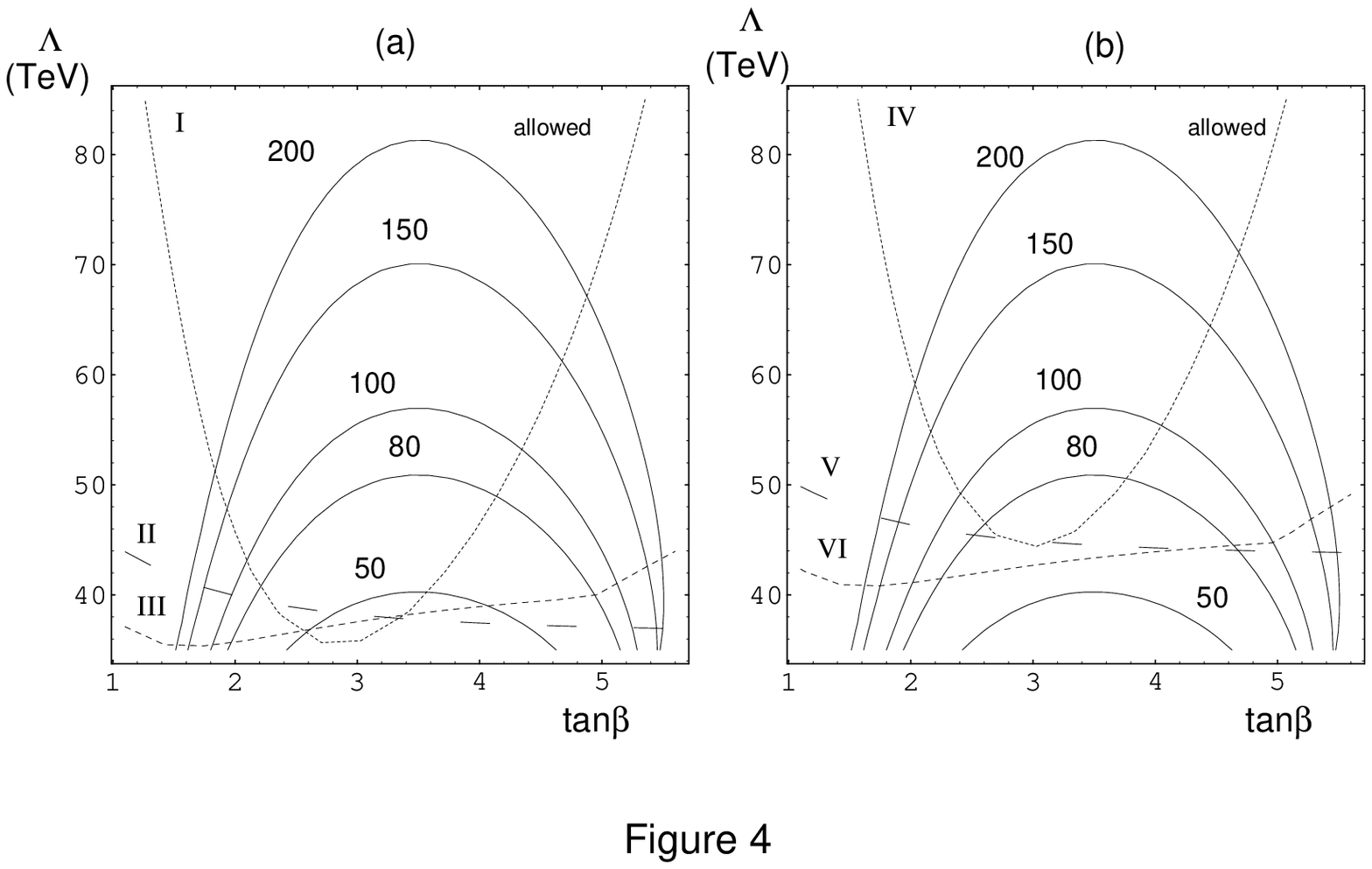}
\protect\label{ftdns2}
\end{figure}


\begin{thebibliography}{99}
\bibitem{susskind}See, for example, L. Susskind, 
{\it Phys. Rev.} {\bf 
D20} (1979) 2619.
\bibitem{susy}For reviews of supersymmetry and
supersymmetry phenomenology, see, for example: 
P. Fayet and S. Ferrara, {\it Phys. Rep.} 
\bf{5}\rm (1977) 249;
H.P. Nilles, {\it Phys. Rep.} \bf{110}\rm (1984) 1;
M.F. Sohnius, {\it Phys. Rep.} \bf{128}\rm (1985) 2;
I. Hinchliffe, {\it Ann. Rev. Nucl. Part. Sci.} 
{\bf 36} (1986) 505.
\bibitem{barbieri1}R. Barbieri and G. Giudice, 
{\it Nucl. Phys.} 
{\bf B306} (1988) 63.
\bibitem{anderson}G. Anderson and D. Casta$\tilde{n}$o,
{\it Phys. Lett.}
 {\bf B347} (1995) 300.
\bibitem{georgi}S. Dimopoulos and H. Georgi, {\it
Nucl. Phys.} {\bf B193} (1981) 150.
\bibitem{affleck}I. Affleck, M. Dine and N. Seiberg, {
\it Nucl. Phys.} {\bf B256} (1985) 557.
\bibitem{hall}L. Hall, J. Lykken and S. Weinberg, 
{\it Phys.
 Rev.} {\bf D27} (1983) 2359.
\bibitem{gabbiani}F. Gabbiani and A. Masiero,
{\it Nucl. Phys.} {\bf B322}
(1989) 235; J. S. Hagelin, S. Kelley and T. Tanaka,
{\it Nucl. Phys.}
{\bf B415} (1994) 293; F. Gabbiani, E. Gabrielli,
A. Masiero and
L. Silvestrini, {\it Nucl. Phys.} {\bf B477}
(1996) 321.
\bibitem{wise}L. Alvarez-Gaume, M. Claudson and M. B. Wise,
{\it Nucl. Phys.} {\bf B207} (1982) 96.
\bibitem{nelson2}M. Dine, A. Nelson, Y. Nir and
Y. Shirman, {\it Phys. Rev.} {\bf D53} (1996) 2658.
\bibitem{arkani}N. Arkani-Hamed,
C. D. Carone, L. J. Hall and H. Murayama,
{\it Phys. Rev.} {\bf D54} (1996) 7032.
\bibitem{strumia}P. Ciafaloni and A. Strumia, 
hep-ph/9611204;
G. Bhattacharyya and
A. Romanino, hep-ph/9611243.
\bibitem{fayet} P. Fayet, {\it Nucl. Phys.} {\bf B90}
(1975) 104.
\bibitem{splitting}H. Georgi and S.L. Glashow, 
{\it Phys. Rev. Lett.}
{\bf 32} (1974) 438.
\bibitem{barbieri2}R. Barbieri, G. Dvali and A. Strumia,
{\it Phys. Lett.}
 {\bf B333} (1994) 79.
\bibitem{nelson}M. Dine, A. Nelson and Y. Shirman,
{\it Phys. Rev.} {\bf D51} (1995) 1362.
\bibitem{randall}I. Dasgupta, B. A. Dobrescu and 
L. Randall,
{\it Nucl. Phys.} {\bf B483} (1997) 95. 
\bibitem{unpublished}S. Dimopoulos, G. Giudice
and A. Pomarol, {\it Phys. Lett.} {\bf B389} (1996) 37;
S. Martin,
{\it Phys. Rev. } {\bf D55} (1997) 3177.
\bibitem{thomas}See for example,
S. Dimopoulos, S. Thomas and J. Wells, hep-ph/9609434 
and references
therein.
\bibitem{aleph}Talk presented by Glen Cowan (ALEPH 
collaboration) at
the special CERN particle physics seminar on physics 
results from the LEP
run at 172 GeV, 25 February, 1997.
\bibitem{cerngroup2}G.F. Giudice, M.L. Mangano, 
G. Ridolfi and
R. Ruckel ({\it convenors}), {\it Searches for 
New Physics},
hep-ph/9602207.
\bibitem{langacker}See, for example, J. Ellis, 
S. Kelley, and
D. V. Nanopoulos, {\it Nucl. Phys.} {\bf B373} (1992) 55;
P. Langacker and N. Polonsky,
{\it Phys. Rev.} {\bf D47} (1993) 4028.
\bibitem{pdg}{\it Review of Particle Physics},
{\it Phys. Rev.} {\bf D54} (1996) 1.
\bibitem{barr}S. M. Barr, hep-ph/9607359.
\bibitem{wilczek}F. Wilczek, {\it Phys. Rev. Lett.}
{\bf 40} (1978) 279.
\bibitem{cerngroup}M. Carena and P. W. Zerwas 
({\it convenors}),
{\it Higgs Physics}, CERN Yellow Report CERN 96-01, 
hep-ph/9602250.
\bibitem{delphi}W. Adam {\it et al}, DELPHI Collaboration,
CERN-PPE/96-119.
\bibitem{srednicki} H.P. Nilles, M. Srednicki and D. Wyler,
{\it Phys. Lett.} {\bf B124} (1983) 337; A.B. Lahanas,
{\it Phys. Lett.} {\bf B124} (1983) 341.

\end{thebibliography}
\end{document}